\documentclass[preprint,review,12pt,leqno]{elsarticle}
\usepackage{float}
\usepackage{graphicx}
\usepackage{grffile}
\usepackage{multirow}
\usepackage{gensymb}
\usepackage{booktabs}
\usepackage{tabularx}
\usepackage{footnote}
\usepackage{amsmath} 
\usepackage{color,amsmath,amssymb}
\usepackage{color}
\usepackage{comment}
\usepackage{amssymb}
\usepackage{subcaption}
\usepackage{tikz}

\newcommand{\etal}{{\it et al.} }

\begin{document}

\title{Calculated line lists for H$_{\text{2}}$$^{\text{16}}$O and H$_{\text{2}}$$^{\text{18}}$O with extensive comparisons to theoretical and experimental sources including the HITRAN2016 database}

\author{Eamon K. Conway\footnote{To whom correspondence should be addressed; email: eamon.conway@cfa.harvard.edu}} 
 \address{Center for Astrophysics $|$ Harvard \& Smithsonian,  Atomic and Molecular Physics Division, Cambridge, MA, USA. 02138}
\address{Department of Physics and Astronomy, University College London, Gower Street, London WC1E 6BT, United Kingdom}

\author{Iouli E. Gordon}
 \address{Center for Astrophysics $|$ Harvard \& Smithsonian,  Atomic and Molecular Physics Division, Cambridge, MA, USA. 02138}

\author{Aleksandra A. Kyuberis}
 \address{Institute of Applied Physics, Russian Academy of Sciences, Ul'yanov Street 46, Nizhny Novgorod, Russia 603950}
 
\author{Oleg L. Polyansky}
\address{Department of Physics and Astronomy, University College London, Gower Street, London WC1E 6BT, United Kingdom}
 \address{Institute of Applied Physics, Russian Academy of Sciences, Ul'yanov Street 46, Nizhny Novgorod, Russia 603950}

\author{Jonathan Tennyson} 
 \address{Department of Physics and Astronomy, University College London, Gower Street, London WC1E 6BT, United Kingdom}
 
\author{ Nikolai F. Zobov}
 \address{Institute of Applied Physics, Russian Academy of Sciences, Ul'yanov Street 46, Nizhny Novgorod, Russia 603950}
 
\begin{abstract}

 New line lists are presented for the two most abundant water  isotopologues; H$_{2}$$^{16}$O and H$_{2}$$^{18}$O. The  H$_{2}$$^{16}$O line list extends to 25710 cm$^{-1}$ with intensity stabilities provided via ratios of calculated intensities obtained from two different semi-empirical potential energy surfaces. The line list for H$_{2}$$^{18}$O extends to 20000 cm$^{-1}$. The  minimum intensity considered for all is $10^{-30}$ cm  molecule$^{-1}$ at 296~K, assuming 100\% abundance for each  isotopologue. Fluctuation of calculated intensities caused by changes in the underlying potential energy are found to be  significant, particularly for weak transitions. Direct comparisons  are made against eighteen different sources of line intensities, both  experimental and theoretical, many of which are used within the  HITRAN2016 database.  With some exceptions, there is excellent agreement between our line lists and the experimental intensities in  HITRAN2016. In the infrared region, many H$_{2}$$^{16}$O bands which exhibit intensity differences of 5-10\% between to the most recent  'POKAZATEL' line list (Polyansky \textit{et al.}, [Mon. Not. Roy. Astron. Soc. \textbf{480}, 2597 (2018)] and observation,  are now generally predicted to within 1\%. For H$_{2}$$^{18}$O, there are systematic differences in the strongest intensities calculated in  this work versus those obtained from semi-empirical calculations. In the visible, computed cross sections show smaller residuals between our work and both HITRAN2016 and HITEMP2010 than POKAZATEL. While our line list accurately reproduces HITEMP2010 cross sections in the observed region, residuals produced from this comparison do however highlight the need to update line positions in the visible  spectrum of HITEMP2010. These line lists will be used to update many transition intensities and line positions in the HITRAN2016 database.

\end{abstract}
\maketitle

\newpage
\section{Introduction}

Water vapor is one of the most well studied and analyzed molecules in existence. Countless experiments (for instance those in Refs. \cite{jt269,H2O-S-15-11,H2O-S-18-15,H2O-S-27-23,H2O-S-28-24,H2O-nu-59-1235,jt687,H2O-S-63-1212,H2O-S-64-1213,H2O-S-65-1214,H2O-S-67-1216,H2O-S-68-1217,H2O-S-69-1218,PTASHNIK201692,Grechko2010,MIKHAILENKO2018170}) and \textit{ab initio}\cite{jt394,ps97,sp00,jt378,jt424,11BuPoZo.H2O,jt734,jt714} calculations have been performed to document its spectrum; such research is presently on-going and is likely continue for many more years. 

Water is abundant in our own atmosphere, where it is a source of significant radiation absorption at all wavelengths, but is readily observed in the spectra obtained from many extra-terrestrial bodies, not limited to but including exoplanets\cite{jt400,Konopacky1398}, comets\cite{comets,jt330} and cool stars\cite{jt173}.

The absorption spectrum of this asymmetric top molecule is complex and
can be observed in the Earth's atmospheres from the microwave to the near
ultraviolet. Water spectra can influence
the retrieval of many molecules. When accurate retrievals are sought,
precise intensities and line parameters for almost every transition
become important. For those molecules with trace abundances, extra
emphasis is placed on this to reduce the unwanted
interference\cite{amt-10-3833-2017,jt645}.

The HITRAN\cite{jt691} database is a widely acknowledged and thoroughly verified source of spectroscopic data, housing line-by-line parameters for 49 molecules that are detectable in our own and some planetary atmospheres. The database is used in a variety of applications, from the calibration of instruments, to the interpretation of telluric spectra and modeling of planetary atmospheres, and as such, the data which makes its way into the database is extensively verified, not just by the HITRAN group, but by the scientific community that extensively uses it.  
The data which forms the H$_{2}$$^{16}$O and H$_{2}$$^{18}$O  line lists in HITRAN2016\cite{jt691} originate from a combination of \textit{ab initio}, semi-empirical and experimental methods.

Experiments can have the advantage over \textit{ab initio} methods by possessing the ability to measure transition parameters to high levels of precision (especially for the line positions), although the scale at which this is accomplished is often limited to one particular spectral region. Theoretical methods can, however, calculate entire spectra. The agreement between observation and calculation can often fluctuate from band to band, a product of both experimental and theoretical limitations, and the different systematic errors which characterize the two techniques. A combination of both resources are necessary to produce an accurate spectrum. Calculated spectra are now capable of predicting energy levels to tenths of a wavenumber, a feature that is essential for interpreting telluric observations. 

The spectroscopically-determined potential energy surface of Bubukina \textit{et al.}\cite{11BuPoZo.H2O} predicts energy levels that are below 26000 cm$^{-1}$ to 0.022 cm$^{-1}$, while the more recent semi-empirical potential of Mizus \textit{et al.}\cite{jt714} calculates levels beneath 15000 cm$^{-1}$ to 0.011 cm$^{-1}$, which approaches the resolution of many atmospheric spectrometers on board satellite missions. 

The most recently available calculated line list for H$_{2}$$^{16}$O, called POKAZATEL is due to Polyansky \textit{et al.}\cite{jt734}. POKAZATEL is the first line list for a tri-atomic that is complete; it includes essentially every possible transition up to rotational level 72, where above J$=72$, the lowest energy level that is populated occurs above dissociation. The POKAZATEL line list used the dipole moment surface (DMS) of Lodi \textit{et al.}\cite{jt509}, which was published in 2011. Since then, many high quality experiments have been conducted, from the infrared to the visible and small imperfections in the DMS have been observed. Sironneau and Hodges \cite{H2O-S-69-1218} measured intensities near 1.25 $\mu$m to high precision, citing a sub percent uncertainty. They showed that there are inconsistencies in the prediction of 2$\nu_{3}$ band intensities. 

{Birk \textit{et al.} \cite{jt687} analyzed five infrared experiments covering 1250 - 1750 cm$^{-1}$\cite{BIRK1250} , 1850 - 2280 cm$^{-1}$\cite{H2O-nu-59-1235}, 2390 - 4000 cm$^{-1}$\cite{H2O-nu-59-1235}, 4190 - 4340 cm$^{-1}$\cite{BIRK4190} and 10000 - 11000 cm$^{-1}$}. For some absorption bands at 1 $\mu$m, the most recent \textit{ab initio} models were found to no longer predict intensities to within 1-2\%.

Lampel and co-workers\cite{jt645} showed that POKAZATEL and other line lists failed to accurately predict atmospheric absorption in the 500 nm - 450 nm interval. The high temperature database, HITEMP2010\cite{HITEMP2010}, modeled the absorption significantly better, despite being almost ten years old at the time the work was carried out. HITEMP2010 used the calculated BT2\cite{jt378} line list as a starting point, with experimental energy levels and intensities from HITRAN2008 replacing theoretical values where possible. 

The motivation behind this work is to create new, highly accurate, calculated line lists for the two most abundant water isotopologues, validated through a significant number of comparisons against high quality experimental data present in the HITRAN2016 database and beyond, as well as semi-empirical methods, covering transition frequencies from the microwave to the visible.

\section{Line List Calculation}

For the HITRAN2016\cite{jt691} H$_{2}$$^{16}$O line list, the maximum rotational level considered is $J=20$, with a cut off transition frequency of 25710 cm$^{-1}$ and our line list follows the same thresholds. For H$_{2}$$^{18}$O this is 20000 cm$^{-1}$. Assuming 100\% abundance for each isotopologue, we choose to calculate transitions with a minimum intensity of 10$^{-30}$ cm molecule$^{-1}$ at the standard room temperature of $T=296$ K which is sufficient for the majority of studies carried out at the range of temperatures encountered in the terrestrial atmosphere. When included in atmospheric models it maybe practical to further truncate the line lists by including isotopic abundance under the same cut off criteria.  In the far infrared of HITRAN2016, there are several transitions with intensities orders of magnitude weaker than this, however, they are negligible in comparison to many pure rotational transitions. In cases where one wants to make use of weaker transitions (for instance in high-temperature applications) the HITEMP database (which is beyond the scope of this paper) should be used instead of HITRAN.

To calculate our line lists, we consider the semi-empirical potential energy surfaces (PES) of Bubukina \textit{et al.}\cite{11BuPoZo.H2O} and Polyansky \textit{et al.}\cite{jt734} (POKAZATEL PES) with the most recent \textit{ab initio} dipole moment surface available, CKAPTEN\cite{jt744}. It is important to distinguish the POKAZATEL line list from the POKAZATEL potential: the POKAZATEL line list used the dipole surface of Lodi \etal \cite{jt509} with the POKAZATEL potential for the calculation of its spectra. 

The high-accuracy IR potential of Mizus \etal \cite{jt714}, termed PES15k, calculates energy levels to an average deviation of only 0.011 cm$^{-1}$ when compared to experiment. However, it only covers energy levels to 15000 cm$^{-1}$, which is too low for this work, hence we do not consider it here. 

We use the DVR3D suite of programs of Tennyson \textit{et al.}\cite{jtDVR} for the generation of all spectra. These semi-empirical PESs began with an \textit{ab initio} potential, which is then fit to model observed spectroscopic data. Our line lists are hence computed with the wavefunctions of a semi-empirical potential and an \textit{ab initio} dipole. 

For H$_{2}$$^{16}$O measured energy levels below 26000 cm$^{-1}$, the PES of Bubukina \textit{et al.} predicts these states to an average deviation of 0.022 cm$^{-1}$, whilst the POKAZATEL PES calculates the same levels to a RMS of 0.04 cm$^{-1}$. The POKAZATEL potential has the added feature that it accurately predicts energy levels above 26000  cm$^{-1}$. For measured $J=0$ levels between 0 cm$^{-1}$ and dissociation at 41 145 cm$^{-1}$ \cite{jt549}, POKAZATEL has a RMS of 0.13 cm$^{-1}$. Bubukina and co-workers also modified their potential for the other isotopologue H$_{2}$$^{18}$O, which we also use in the calculation of spectral intensities.

To obtain a measure or indication of the theoretical uncertainty on transition intensities, the method of Lodi and Tennyson\cite{jt522} involves computing spectra with two different dipoles and two different potentials. The uncertainty of each line is assessed by taking the largest ratio between the transition intensities obtained from each of the four possible calculations. A ratio close to unity would indicate a stable transition, where a ratio deviating from one indicates that the transition is 'sensitive' to either the DMS or PES. 

For unstable weak transitions, it is understood that the underlying potential is the cause of such instability, while systematic errors in band intensities that may appear for stable lines are likely due to the DMS . To quantify an uncertainty, one requires knowledge of both factors. For H$_{2}$$^{16}$O, we aim, in this work, to identify unstable transitions which are solely due to the potential. Hence, we will consider one DMS and two different potentials. Also, the POKAZATEL line-list used the second most recent DMS that is available, created by Lodi \etal \cite{jt509} in 2011, and we will be comparing to the POKAZATEL line-list. 

The dipole calculations for this DMS were carried out at a very high level of theory, where each point ($\approx$16000 total) required over 140000 seconds of CPU time to complete\cite{jt744}. For high-accuracy theoretical spectra, the Multi-Reference Configuration Interaction (MRCI) method combined with the aug-cc-pCV6Z basis set is the current gold-standard formalism, however, it is limited to only the lightest of molecules with few degrees of freedom, such as H$_{3}$$^{+}$ \cite{FCIh3+} and HCN \cite{jt689}. Hence, we expect the DMS to provide excellent transition intensities for strong, stable lines in the IR, where calculated intensities will deviate by no more than 1-2\% from high-accuracy experiments. For stable transitions in the visible/near-UV, the accuracy of both experiment and theory is expected to reduce as the majority of these will be weak, however, the agreement should still be within 10-20\%. Through a significant number of comparisons, against both experimental and theoretical sources, the quality of this DMS, CKAPTEN, will be verified. Where we quote theoretical uncertainty below, we refer to the instability induced by the potential energy surface.

For H$_{2}$$^{16}$O, using the POKAZATEL PES with the CKAPTEN DMS, we created a temporary line list that extended slightly beyond the HITRAN2016 cut-off st 25000 cm$^{-1}$. and went to 26000 cm$^{-1}$. Similarly, using the PES of Bubukina \textit{et al.} with CKAPTEN we created another line list that also extends to 26000 cm$^{-1}$. Two line lists are required to obtain stabilities.

The POKAZATEL PES was refined using  the same number of measured energy levels (below 26000 cm$^{-1}$) to that of Bubukina \textit{et al.} within rotational levels J = 0, 2, 5. Since the PES of Bubukina \textit{et al.} provides more accurate energy levels, we use the line list that included this PES with the CKAPTEN DMS for our main/reported line list, with the other line list only used for the purpose of obtaining a PES stability on the intensities. 

The H$_{2}$$^{18}$O  line list  use the PES of Bubukina \textit{et al.} with the CKAPTEN DMS.

\subsection{MARVELization}

MARVEL (measured active rotational-vibrational energy levels)
\cite{jt412} is a procedure for inverting observed spectroscopic
frequencies to obtain well characterized empirical energy levels and
uncertainties. The method was developed for a task group studying the
spectroscopy of water \cite{jt562} who provided energy levels for all
the isotopologues of water \cite{jt454,jt482,jt539,jt576}. As
illustrated by this task group \cite{jt482} use of these empirical
levels facilitates not only the reproduction of the original observed
line position but also the precise prediction of many other
yet-to-be-observed line positions between states for which there are
MARVEL energy levels. The MARVEL algorithm has been systematically
improved \cite{12FuCsa,jt750} with a particular focus on continuing to
improve the energy levels of water \cite{jt750,08FuCs,jtwaterupdate}.
An important point is that MARVEL requires all transitions, and hence
resulting energy levels, to be fully labeled.

DVR3D provides rigorous quantum numbers, notably, an ortho/para label,
rotational quantum number and parity. Where possible, it is helpful to
also label each H$_{2}$$^{16}$O and H$_{2}$$^{18}$O energy level with
quantum numbers $K_{a}$, $K_{c}$, $\nu_{1}$, $\nu_{2}$ and $\nu_{3}$.
On their own, the rigorous quantum numbers from DVR3D do not provide
enough information to match with the empirical energy levels from
MARVEL. Hence, energy level differences must be used together with
this information to match with MARVEL states. This approach makes the
labeling of closely lying states not straightforward.

For H$_{2}$$^{18}$O we supplemented the MARVEL data with new pseudo-experimental
levels taken from Polyansky \textit{et al.} \cite{jt665}.

 Each energy level calculated with DVR3D is unique, but can naturally occur more
than once as both an upper or lower state within a transition. It is
therefore important that the same energy level from DVR3D is not
assigned differently in various transitions.
We approached this matching problem in stages, where the maximum energy difference interval for matching, i.e   $|E_{calc}\;-\;E_{meas}|$, is increased in values of $0.22$ cm$^{-1}$, $0.40$ cm$^{-1}$, $0.50$ cm$^{-1}$, $0.70$ cm$^{-1}$, $1.00$ cm$^{-1}$, $1.30$ cm$^{-1}$ and $1.50$ cm$^{-1}$. Increasing beyond this final value would dramatically increase the possibility of mis-labeling.

Using the uniqueness, once a calculated level from DVR3D is matched to its corresponding level in MARVEL, only that particular calculated level may carry that MARVEL label from then on. Also, once a MARVEL level is matched to a level in our line list, it does not carry onto the next iteration, thus preventing duplication from ever occurring. 

For our line lists, where matching was successful, our calculated energy levels were replaced with what is predicted from MARVEL. Those levels that were not assigned in the final iteration were given a default/unknown label of ($\nu_{1}$, $\nu_{2}$, $\nu_{3}$, $K_{a}$, $K_{c}$) = $-2$. We note that there are strong
theoretical grounds for believing that it is not actually possible to label all the states of water
with consistent set of harmonic oscillator, rigid rotor labels above the barrier to linearity \cite{jt234}
which lies at about 11110 cm$^{-1}$ \cite{jt362}.

\subsection{Calculation of H$_{2}$$^{16}$O Stabilities}

Using both H$_{2}$$^{16}$O line lists, we assess the stability in each line intensity based on the ratio of matched intensities \cite{jt522}. Applying this process was not straightforward, as achieving a consistent set of labels for all levels is difficult. For unstable, often weak transitions, the labeling of a particular state may fluctuate between potentials. 

We opted to include an intensity ratio threshold in this matching process, whereby, if transitions in each of our line lists are matched on all quantum numbers and via a small energy level interval, if the ratio of their intensities falls outside the interval $0.66\le S_{1}/S_{2}\le1.5$, then the upper level has its label  reset to default/unknown:  As previously mentioned, this generally occurs for very weak transitions. This method also reduces the possibility of our H$_{2}$$^{16}$O line list having incorrectly labelled states.

\begin{figure}[h]
\centering
\includegraphics[width=0.5\textwidth]{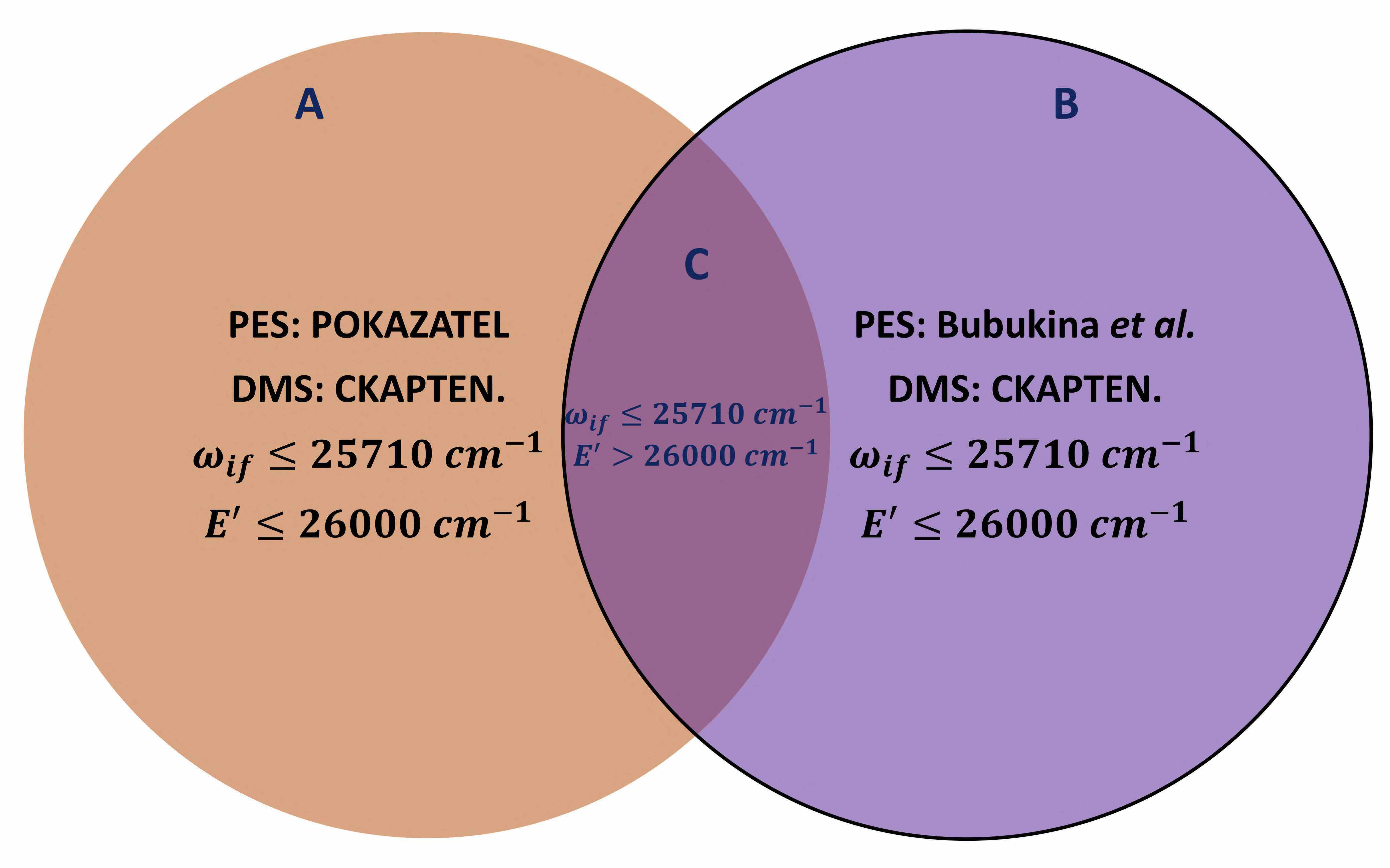}
\caption{$A$ represents the secondary line list computed using the POKAZATEL PES and CKAPTEN DMS. $B$ is our main line list calculated with the Bubukina PES and the CKAPTEN DMS. $C$ are those transitions that have intensities below 10$^{-30}$ cm molecule$^{-1}$, frequencies below 25710 cm$^{-1}$ and E$^{'}$ above 26000 cm$^{-1}$.}
\label{fig:1}
\end{figure}

The Bubukina \textit{et al.} potential is not reliable for accurately calculating energy levels that lie above 26000 cm$^{-1}$, however, there are a small number of transitions with intensities above 10$^{-30}$ cm molecule$^{-1}$ and frequencies below 25710 cm$^{-1}$, that have an upper state energy above 26000 cm$^{-1}$. For these few particular transitions, we replace all the parameters with those produced from the POKAZATEL potential. In Figure \ref{fig:1}, these few transitions are in $A \cup B$, and our final line list is within the solid circle enclosing regions B and C.

\section{Results}
\subsection{H$_{2}$$^{16}$O}

In Table \ref{table:1} we summarize the PES stability percentage on transition intensities which is calculated as $100|(1- \frac{S_{Bubukina~PES}}{S_{POKAZATEL~PES}})|$, and follow the HITRAN2016 relative uncertainty groups. The majority of transitions are stable with PES sensitivity less than 1\%, while the number of unstable transitions appear to decrease as sensitivity increases, an attractive quality in this line list. Those transitions that could not be assigned a sensitivity are not included in the table. 

 \begin{table}[h!]
	\centering
 	\caption{The sensitivity of transition intensities on the underlying potential energy surfaces of Bubukina \etal\cite{11BuPoZo.H2O}  and POKAZATEL\cite{jt734} calculated as $100|(1- \frac{S_{Bubukina}}{S_{POKAZATEL}})|$. The CKAPTEN DMS \cite{jt744} was used for both calculations. }

\begin{tabular}{cc}
	Percentage Range. & Number of Lines.  \\
		\hline 
		\hline 
		 $>$20 \% &4008  \\
		 10~-~20 \% & 5677\\
		 5~-~10 \% &13055\\
			 2~-~5 \% &40364\\
			 1~-~2 \% &34567\\
		 0~-~1 \% &96117\\
\hline
	\end{tabular}
	
				\label{table:1}	
\end{table}

Starting from data contained in the HITRAN2016 database, we compare with over 20000 measured transitions that originate from twelve different experimental sources $\;$
 \cite{ H2O-S-15-11,H2O-S-18-15,H2O-S-27-23,H2O-S-28-24,H2O-nu-59-1235,jt687,H2O-S-63-1212,H2O-S-64-1213,H2O-S-65-1214,H2O-S-67-1216,H2O-S-68-1217,H2O-S-69-1218}. It is important to note, that for all our comparisons to HITRAN2016 data, which are shown below, the data obtained from each source may not be 100\% complete, meaning, not all of the reported data from that particular source features in HITRAN2016. 
 
For microwave spectra, transitions in HITRAN2016 come from \textit{ab initio}\cite{jt509} and semi-empirical calculations\cite{H2O-S-21-31,H2O-S-23-33}. This region is dominated by strong, pure rotational transitions and as such, both the DMS and PES are very well defined. Different \textit{ab initio} models \cite{jt509,jt378} hence exhibit very similar behavior due to the similarity of the underlying \textit{ab initio} calculations in such a well defined region, hence it is necessary to compare with the latest semi-empirical methods that do not use the same formalism as we do. Comparisons are made with the line list taken from the more recent work of Coudert \textit{et al.}\cite{COUDERT201436}, which is used in GEISA2015 \cite{jt636}  
 
Several experimental sources\cite{BIRK1250, BIRK4190,H2O-nu-59-1235} in HITRAN2016 covers the mid-infrared region from 1250 - 4390 cm$^{-1}$ and we thought it necessary to supplement this by a comparison to the experimental data of Ptashnik \textit{et al.}\cite{PTASHNIK201692}, which is not in the database. For the NIR, seven experiments\cite{H2O-S-18-15,H2O-S-63-1212,H2O-S-64-1213,H2O-S-65-1214,H2O-S-67-1216,H2O-S-68-1217,H2O-S-69-1218} provide measurements in the interval 7000 - 8339 cm$^{-1}$, while three experiments \cite{H2O-S-27-23,jt687,H2O-S-15-11} cover 10000 - 14500 cm$^{-1}$. Only one experimental source\cite{H2O-S-28-24} provides data on the remainder of the visible spectrum.  

For all figures shown below, where transition intensities in units of cm/molecule are plotted on the x-axis, these intensities represent our calculated values. 
 
 \begin{figure}[ht]
 \centering
	\includegraphics[width=1.0\linewidth]{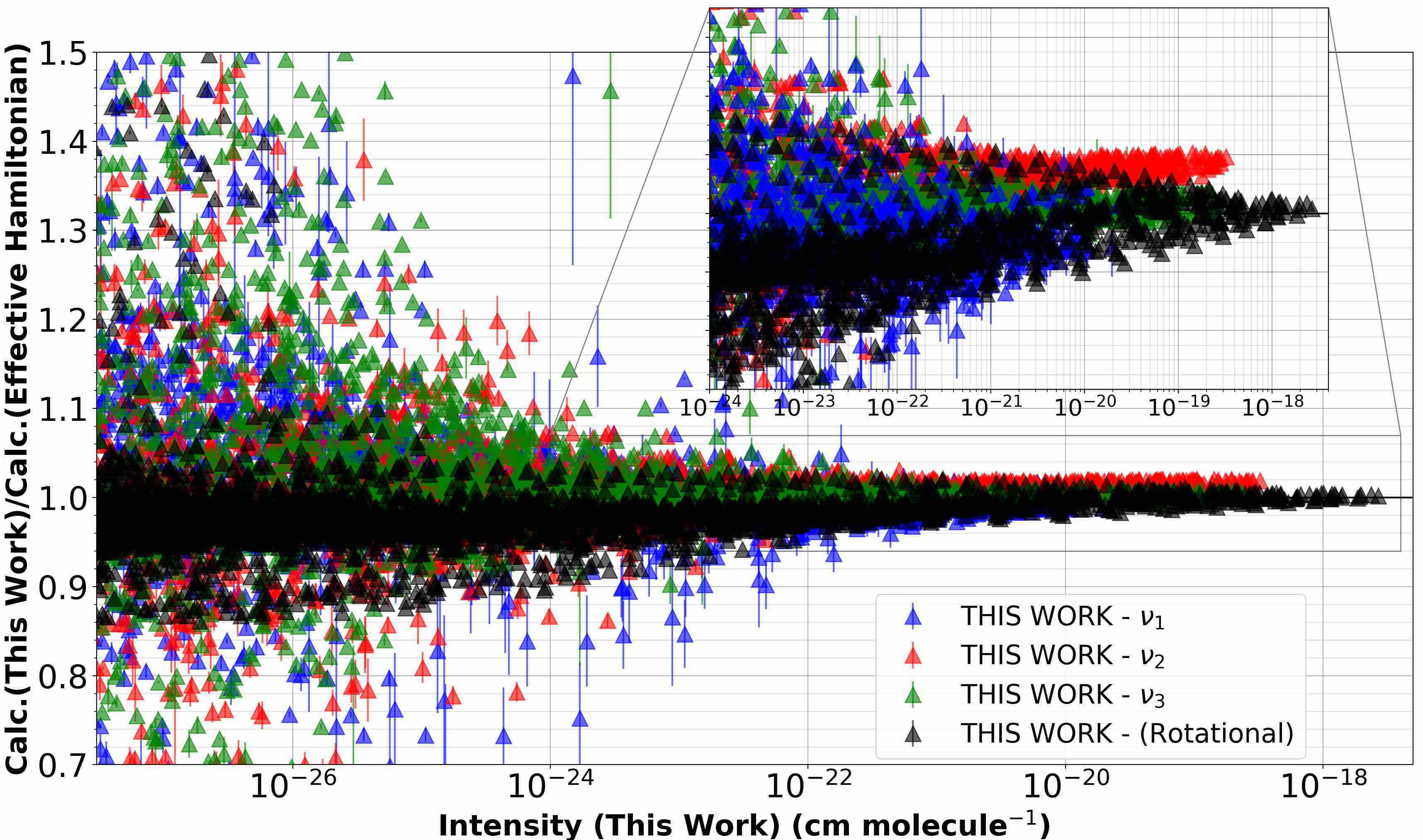}
	\caption{Comparison of our pure rotational, $\nu_{3}$ and $\nu_{2}$ calculated intensities to those from Coudert \textit{et al.}\cite{COUDERT201436}. Uncertainty on our work represents the PES stability.}
	\label{fig:2}
\end{figure}

We compare to 11029 semi-empirical intensities taken from Coudert \textit{et al.}, which include $\nu_{2}$, $\nu_{3}$ and pure rotational transitions. Many of these calculations form the basis of the microwave and far-infrared region in GEISA2015. For the strongest lines above 10$^{-22}$ cm molecule$^{-1}$, the agreement is in general excellent, with the exception of our $\nu_{2}$ intensities being approximately 1\% stronger, a feature that is replicated through recent experiments, seen below in Figure \ref{fig:3}. However, below this threshold, the deviation exhibits a clear rotational dependence. It would appear that this dependence is induced from the work of Coudert \textit{et al.}, as it is not observed in any other of the comparisons below which include the same bands. 
Such behavior has previously been shown to be the consequence of using effective Hamiltonians to represent
intensities \cite{jt625}.

{Figure \ref{fig:3a}  compares 4993 transitions provided in the study of Birk \textit{et al.}\cite{jt687}, which contains measurements from several experiments in the regions of  1250 - 1750 cm$^{-1}$\cite{BIRK1250}, 1850 - 2280 cm$^{-1}$\cite{H2O-nu-59-1235}, 2390 - 4000 cm$^{-1}$\cite{H2O-nu-59-1235} and 4190 - 4340 cm$^{-1}$\cite{BIRK4190}. The agreement is within an average of 1-2\% percent for both our line list and POKAZATEL, although only for those transitions that have intensities over 10$^{-23}$ cm molecule$^{-1}$; below this threshold, the ratios begin to diverge and line instability begins to increase while the accuracy of the experiment begins to decrease.

\begin{figure}
    \centering
    \begin{subfigure}{0.50\textwidth}
        \includegraphics[width=\textwidth]{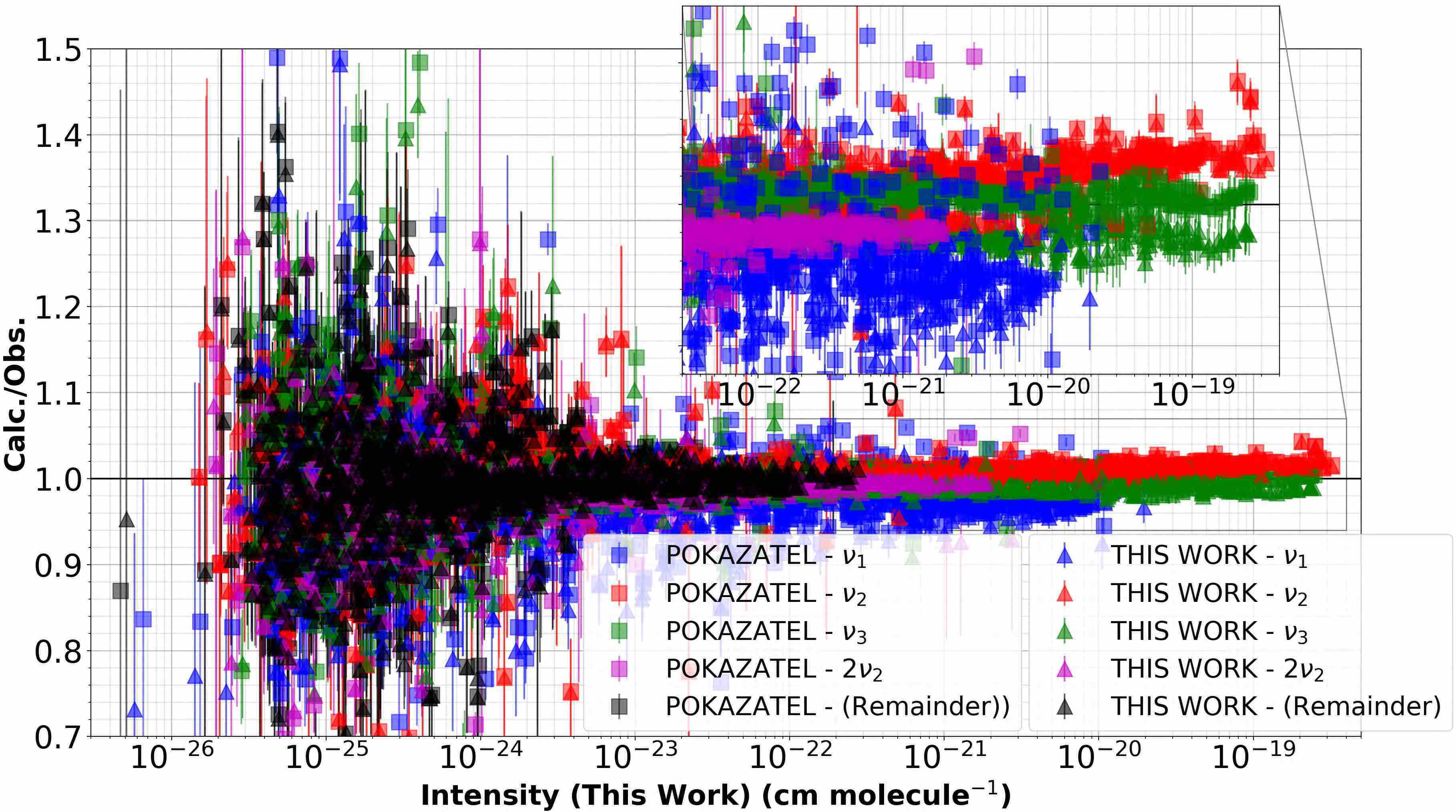}
        \caption{}
        \label{fig:3a}
    \end{subfigure}
    \\
    \centering
    \begin{subfigure}{0.5\textwidth}
        \includegraphics[width=\textwidth]{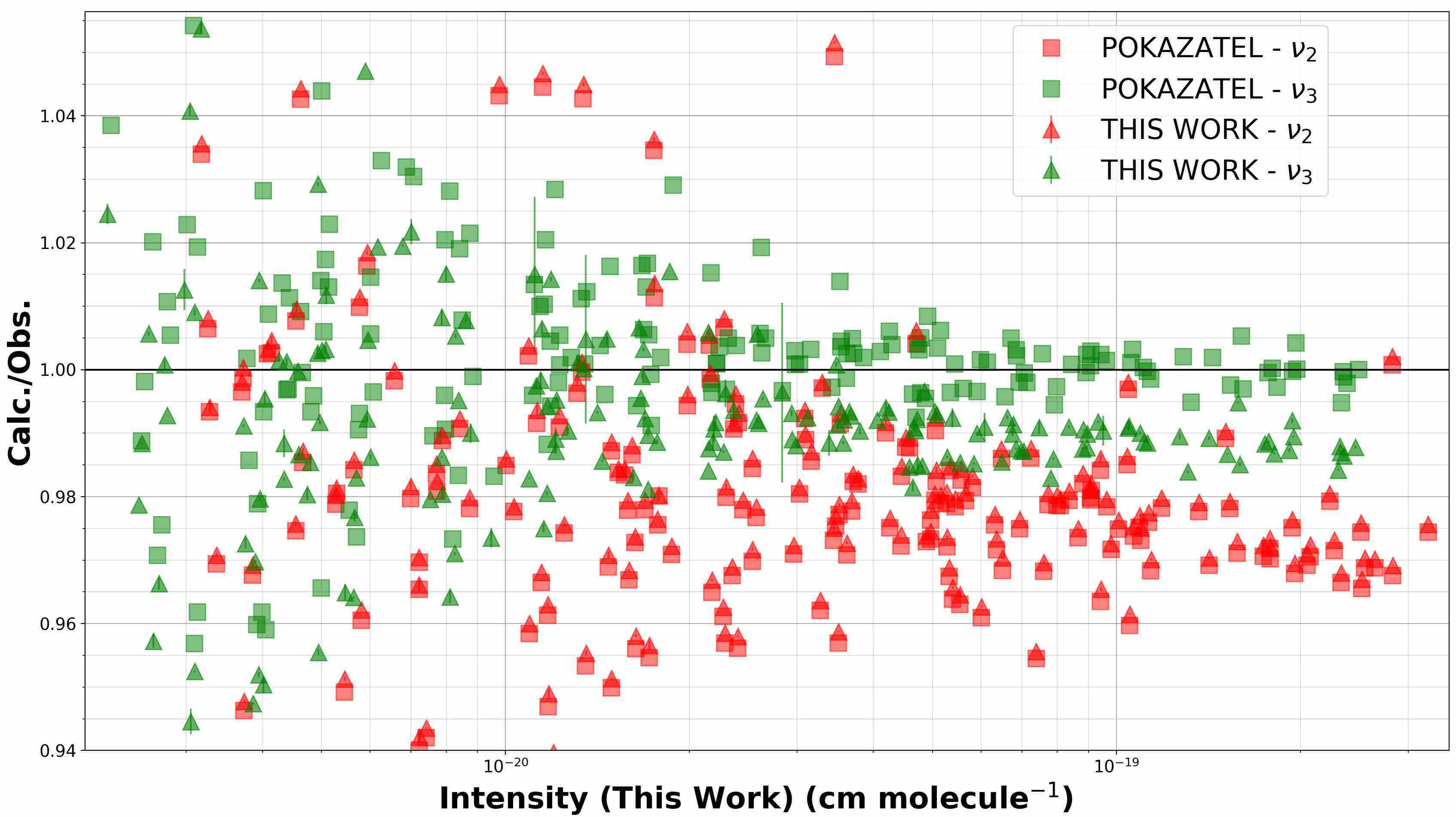}
        \caption{}
        \label{fig:3b}
    \end{subfigure}
    
    \caption{Comparisons of transition intensities from both our new line list and POKAZATEL\cite{jt734} to the measurements of (a) Loos \textit{et al.}\cite{H2O-nu-59-1235}, Birk \etal \cite{BIRK1250,BIRK4190} and (b) Ptashnik \textit{et al.}\cite{PTASHNIK201692}. Uncertainties on this work for (a) are a combination of PES stability and experiment, while POKAZATEL ratios shown in (a) only carry experimental uncertainty. In (b), uncertainty on our data is theoretical stability on the underlying potential.}
    \label{fig:3}
\end{figure}

In the same figure, we show a zoomed in region of the strongest transitions. Ratios of $\nu_{3}$ and $\nu_{2}$ band intensities are in excellent agreement with experiment\cite{H2O-nu-59-1235}. For the $\nu_{1}$ band, both line lists calculate intensities which are approximately 2-3\% too weak, a figure that has already been observed in the study of Birk \etal \cite{jt687}. The deviation in this band is likely an issue that originates from the underlying \textit{ab initio} calculations of the dipole surfaces used in this work and POKAZATEL. We continue to investigate this problem. Intensity measurements of the $\nu_{3}$ fundamental which were recently made by Devi \etal \cite{MALATHYDEVI201813} deviated from those of Loos \etal by 8-10\%. Our results verify the experiment of Loos \etal

The experiment of Ptashnik \textit{et al.} covers regions 1400 - 1840 cm$^{-1}$ and 3440 - 3970 cm$^{-1}$. We plot the resulting intensity ratios in Figure \ref{fig:3b} and this includes 438 data points. The authors converted their results to the customary HITRAN '.par' file format and intensities were given an error code of either four or five, i.e 5 - 10\% or 10 - 30\% respectively. In Figure \ref{fig:3b}, the $\nu_{3}$ band shows excellent agreement to POKAZATEL, where ours is weaker by approximately 1\%, which is well within experimental uncertainty and is in line with what is obtained from Ref. \cite{H2O-nu-59-1235,BIRK1250,BIRK4190}. However, for the $\nu_{2}$ band, both theoretical line lists predict similar intensities again, but are now  weaker by 3\%. The experimental uncertainty on the data of Ptashnik \textit{et al.} is likely the cause of such discrepancy.

Comparing to the HITRAN2016 database, we match to 793 transitions that originate from the work of Toth \textit{et al.}\cite{H2O-S-18-15}, see Figure \ref{fig:4a}. In the Toth \textit{et al.} H$_{2}$O line list, available through the NASA JPL website;\newline \textit{ http://mark4sun.jpl.nasa.gov/h2o.html}. Reported intensities are a combination of experimental measurements and semi-empirical calculations obtained from fitting a model to observation. The region we compare to in HITRAN2016 contains Toth \textit{et al.} measured intensities, not semi-empirical calculations. In general, the agreement to Toth \textit{et al.} using our line list and POKAZATEL for the strongest lines is good, with deviations averaging approximately 1-2\% and is within the 5\% experimental uncertainty. There does appear to be an un-physical shape to the ratios shown in Figure \ref{fig:4a}, which is associated with the bands (101) and (200), and is unlikely to arise from theory.

\begin{figure}[h!]
    \centering
    \begin{subfigure}{0.50\textwidth}
        \includegraphics[width=\textwidth]{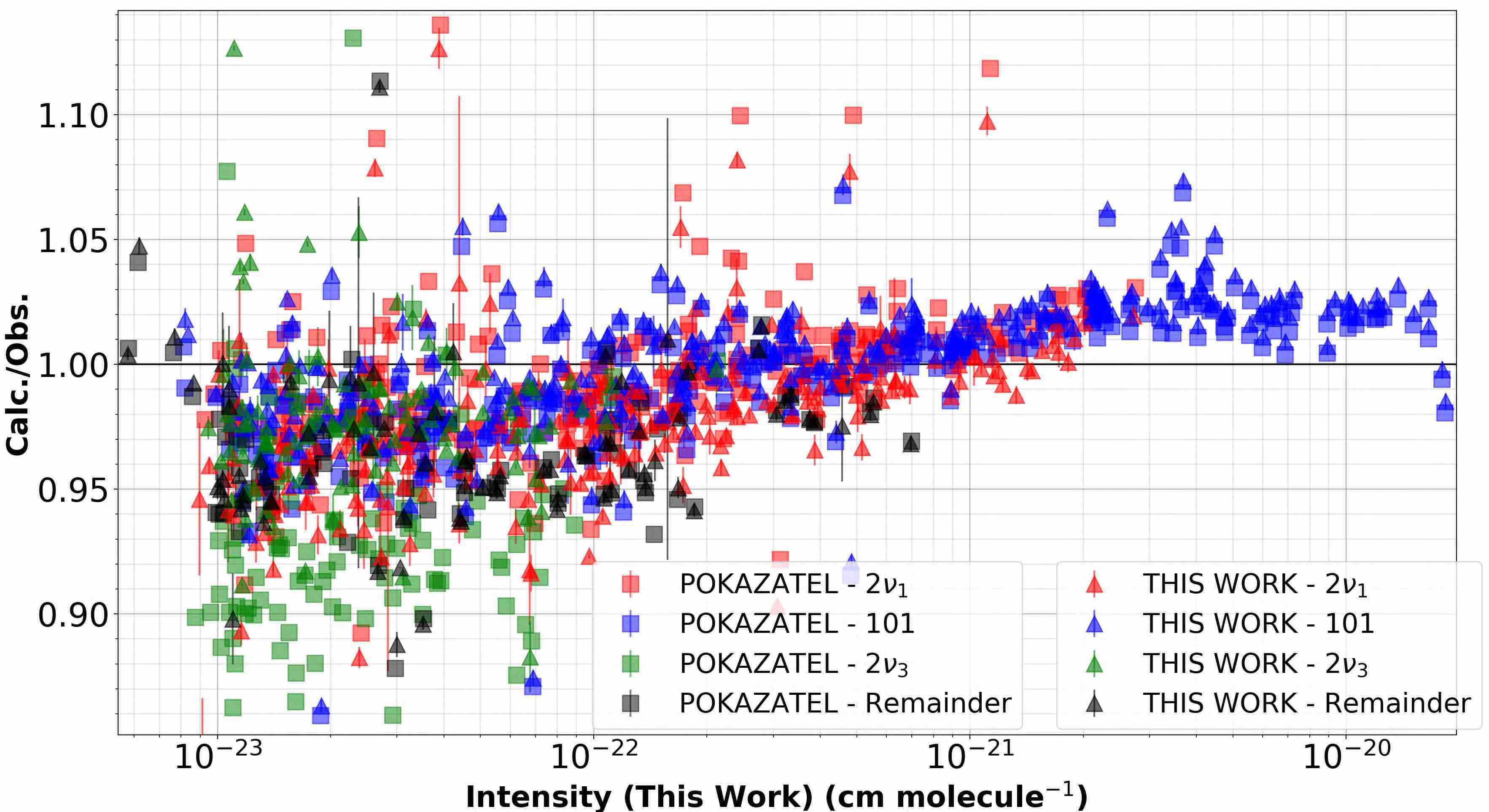}
        \caption{}
        \label{fig:4a}
    \end{subfigure}
    \\
    \begin{subfigure}{0.50\textwidth}
        \includegraphics[width=\textwidth]{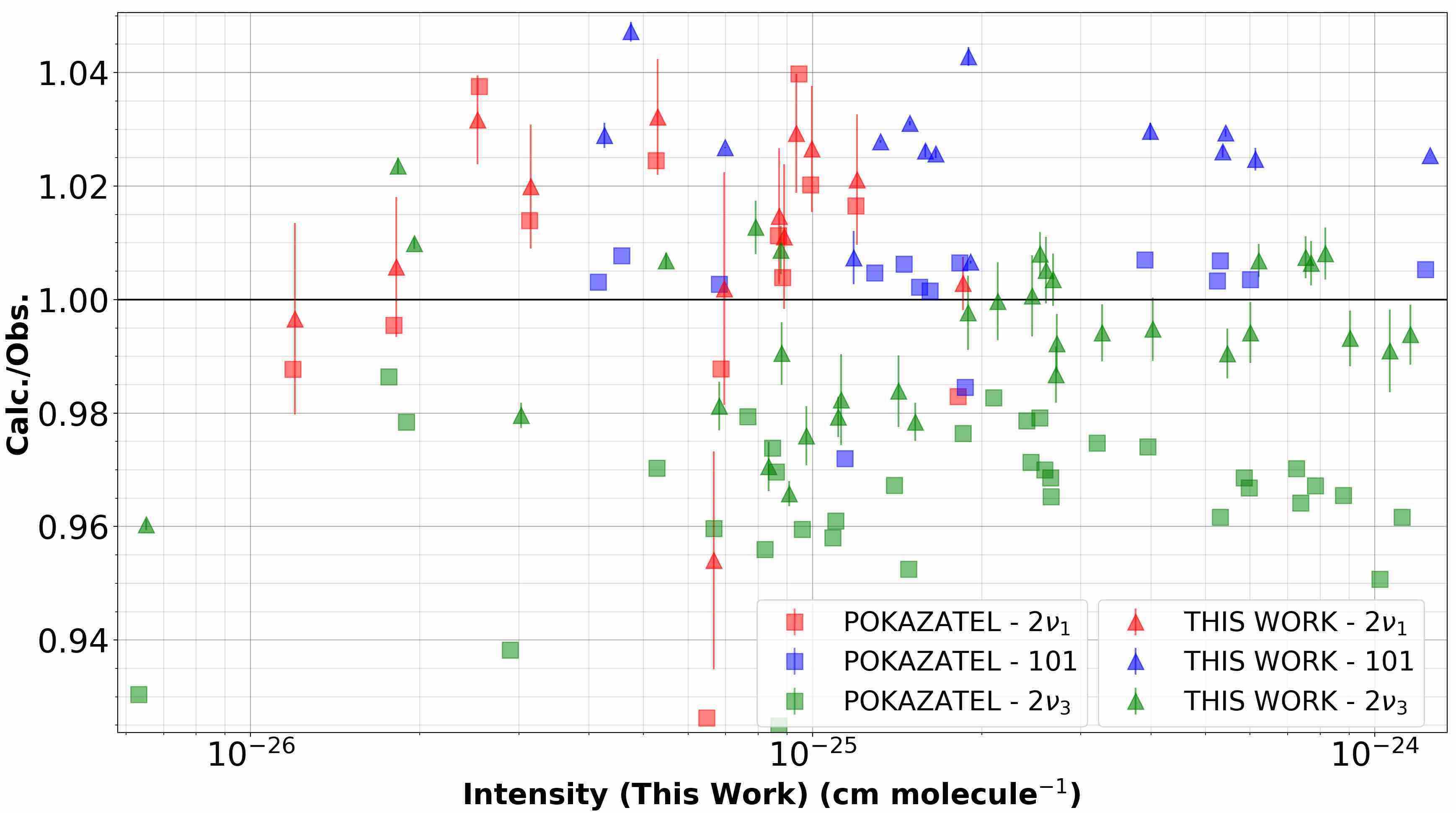}
        \caption{}
        \label{fig:4b}
    \end{subfigure}

    \caption{Intensity ratios from our new line list and POKAZATEL\cite{jt734} to the experimental measurements of (a) Toth \textit{et al.}\cite{H2O-S-18-15} and (b)  Sironneau and Hodges\cite{H2O-S-69-1218}. The uncertainty on our work is calculated from PES stability factor.}
    \label{fig:4}
\end{figure}

Lisak, Havey and Hodges \cite{H2O-S-63-1212} measured intensities of bands (101), (200) and (002) in the narrow wavelength of 7170 - 7183 cm$^{-1}$ to sub-percent accuracy. Both ab initio line lists are in excellent agreement with the fourteen lines from \cite{H2O-S-63-1212} present in HITRAN2016. We chose to omit this figure as there are only a few data points.

Mikhailenko \textit{et al.} \cite{H2O-S-64-1213} measured a large number of weak intensities in the 7408 cm$^{-1}$ - 7919 cm$^{-1}$ window. When comparing to HITRAN2016, we matched to a total of 1227 transitions, with resulting intensity ratios presented in Figure \ref{fig:5a}. Of the two line lists, ours and POKAZATEL, neither shows better agreement to Mikhailenko \textit{et al.} than the other; both theoretical models have scattered ratios throughout.

\begin{figure*}[h!]
    \centering
    \begin{subfigure}{0.45\textwidth}
        \includegraphics[width=\textwidth]{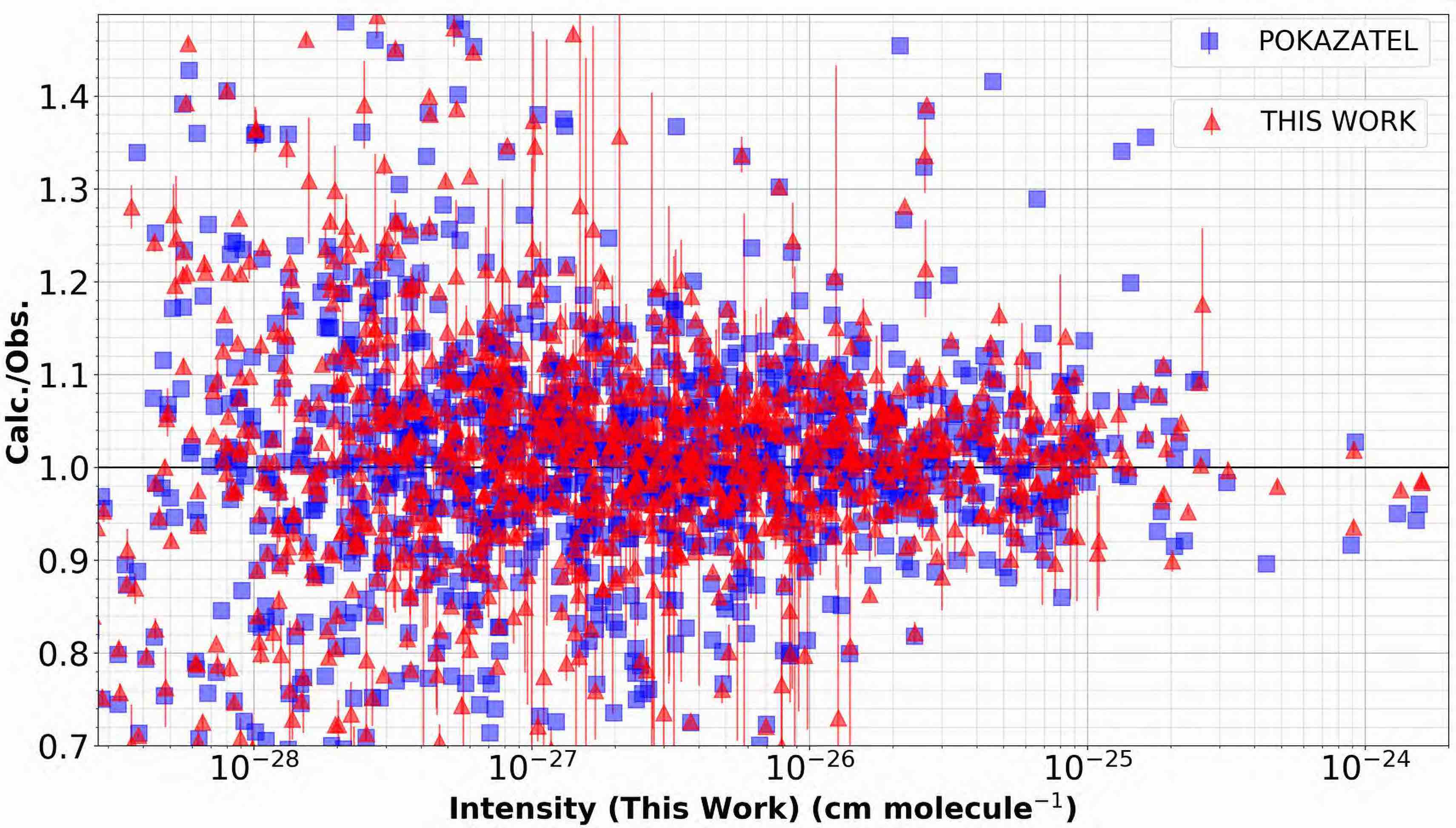}
        \caption{}
        \label{fig:5a}
    \end{subfigure}
    ~
    \begin{subfigure}{0.45\textwidth}
        \includegraphics[width=\textwidth]{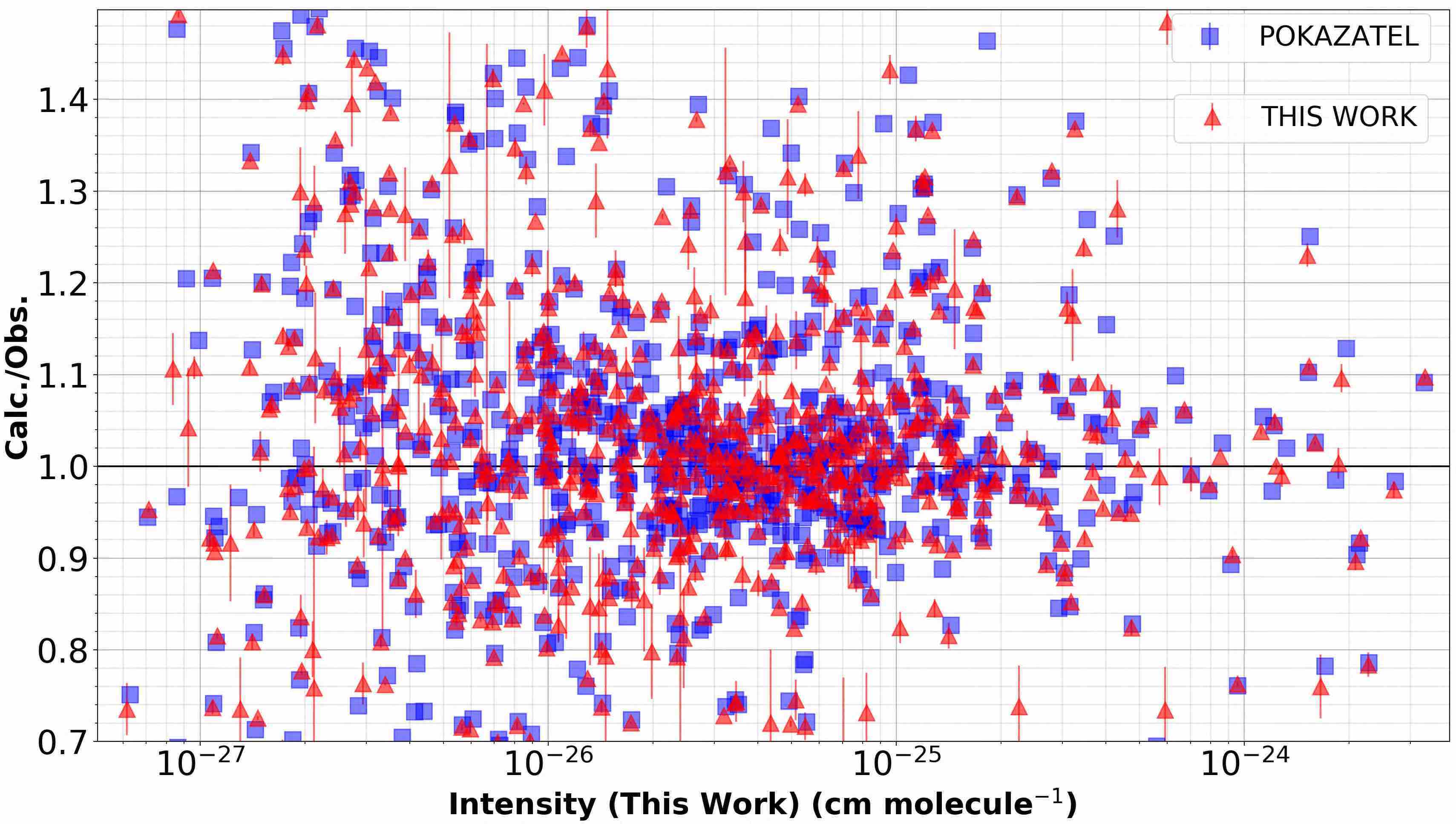}
        \caption{}
        \label{fig:5b}
    \end{subfigure}
\\
    \begin{subfigure}{0.45\textwidth}
        \includegraphics[width=\textwidth]{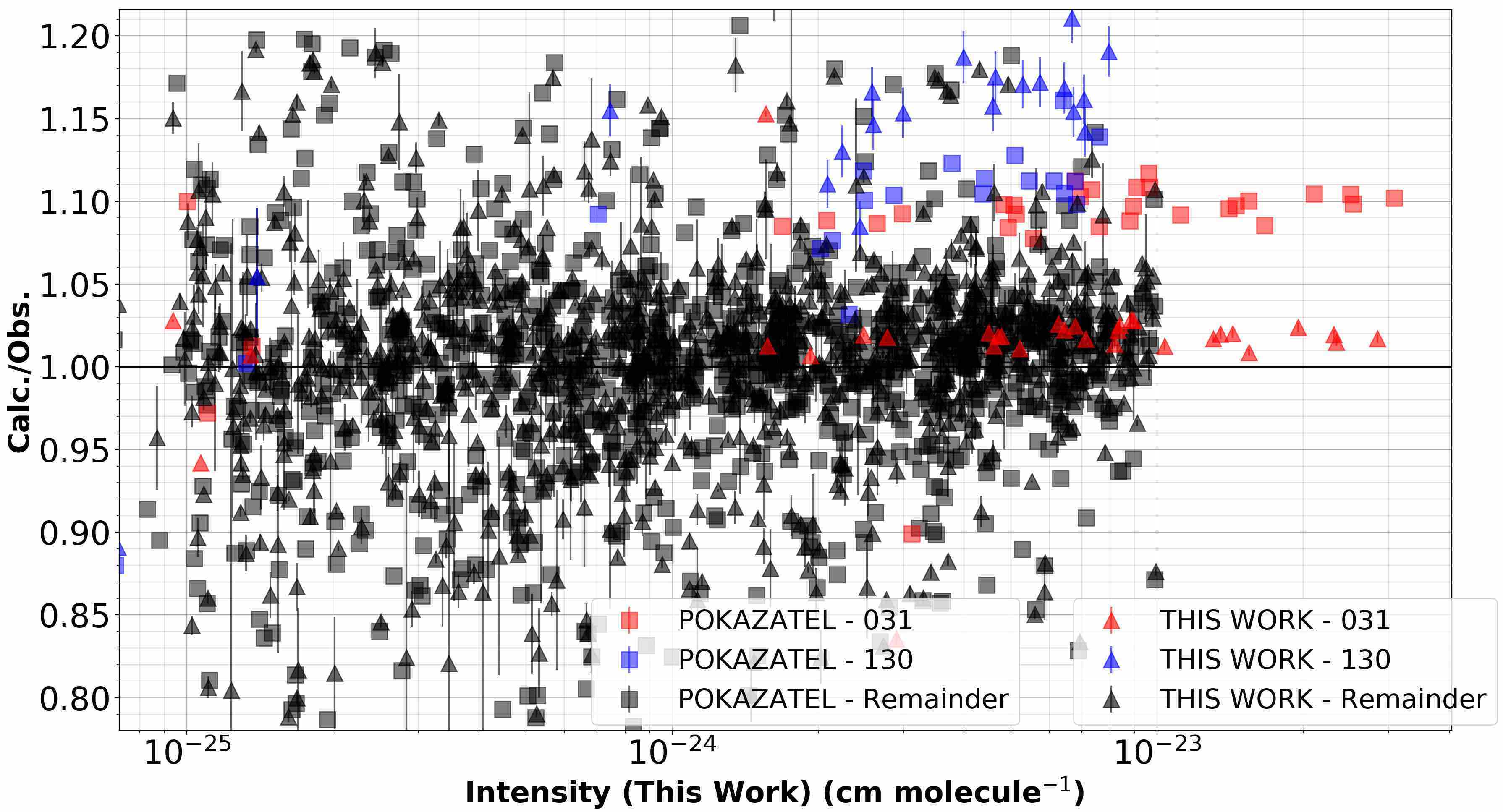}
        \caption{}
        \label{fig:5c}
    \end{subfigure}
    ~
    \begin{subfigure}{0.45\textwidth}
        \includegraphics[width=\textwidth]{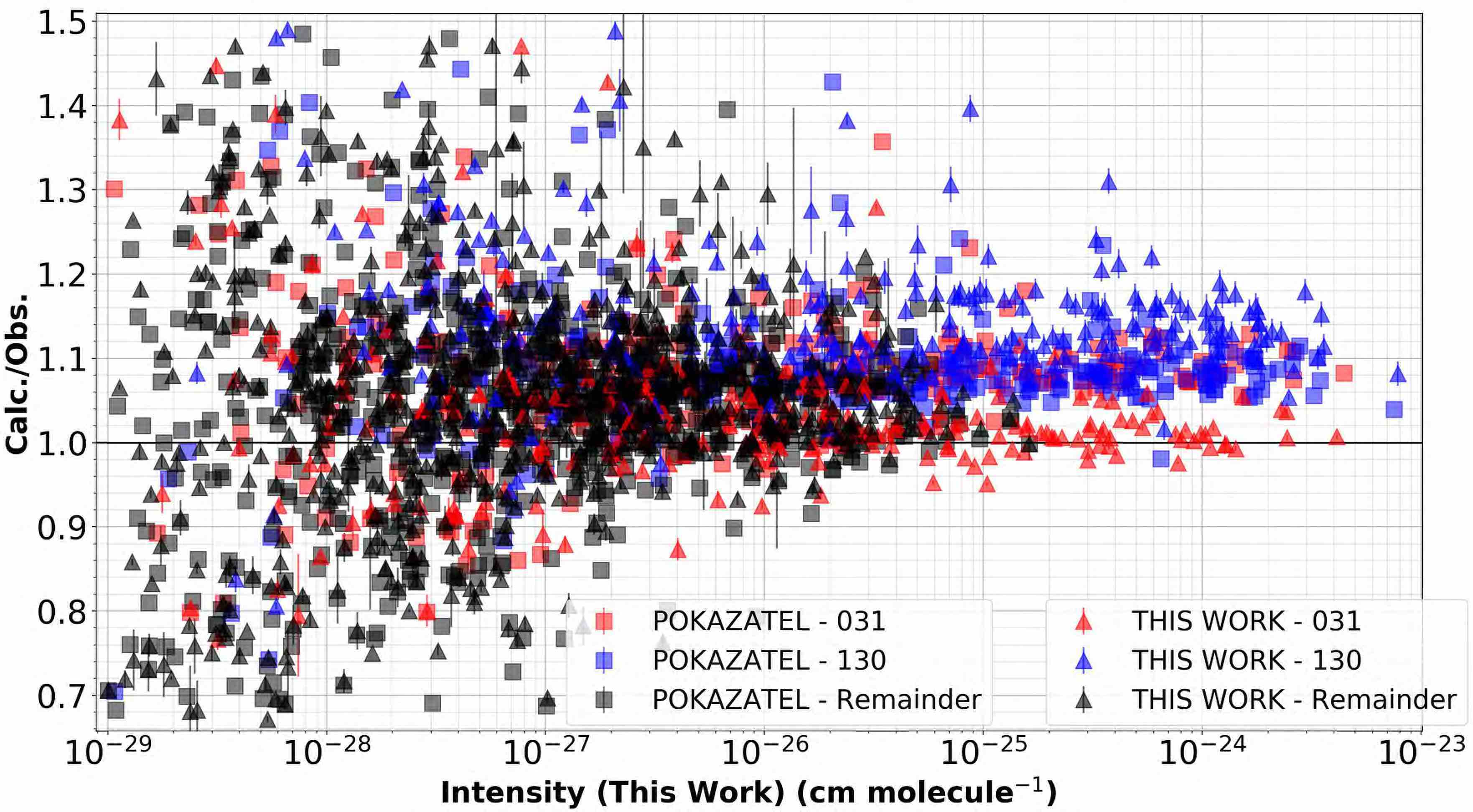}
        \caption{}
        \label{fig:5d}
    \end{subfigure}

    \caption{Comparison of the calculated intensities in our new line list and POKAZATEL\cite{jt734} to the experimental data of (a) Mikhailenko \textit{et al.}\cite{H2O-S-64-1213}, (b) Leshchishina \textit{et al.}\cite{H2O-S-63-1212}, (c) Regalia \textit{et al.}\cite{H2O-S-67-1216} and (d) Campargue \textit{et al}\cite{H2O-S-68-1217} that are all in HITRAN2016. Error bars on our work are from the PES stability factors.}
    \label{fig:5}
\end{figure*}

Leshchishina \textit{et al.}\cite{H2O-S-63-1212} measured transition intensities within a narrow spectral region of 7000 - 7405 cm$^{-1}$. Of the 775 lines shown in Figure \ref{fig:5b}, neither line list out-performs the other and ratios are again, scattered.

Our new line list exhibits excellent agreement with the measurements of Regalia \textit{et al.} \cite{H2O-S-67-1216}, see Figure \ref{fig:5c}. Comparing to HITRAN2016, we matched to 1102 of their measured intensities that cover 7000 - 8339 cm$^{-1}$. Figure \ref{fig:5c} shows a clear discrepancy in the prediction of (031) band intensities of the POKAZATEL line list with a 10\% shift present for the strongest lines. For the (130) band intensities, our calculations are systematically offset by approximately 15-17\%.

In Figure \ref{fig:5d} shows calculated intensity ratios with 1209 measurement from Campargue \textit{et al.}\cite{H2O-S-68-1217}, taken from HITRAN2016, with frequencies in the range 7911 - 8332 cm$^{-1}$. Our line list again shows a large, 10\% improvement over POKAZATEL for the strong intensities of the (031) band. As previously seen in our comparison to the experiment of Regalia \textit{et al.}, our prediction of (130) band intensities are again over-estimated by approximately 14-17\%. The intensity ratios from both experiments exhibit similar results for bands (031) and (130).

Sironneau and Hodges\cite{H2O-S-69-1218} measured intensities with high precision in the 7714 - 7919 cm$^{-1}$ interval. We compare with 65 of their lines that are obtained from our comparison to the HITRAN2016 database; results shown in Figure \ref{fig:4b}. Intensity ratios in the 2$\nu_3$ band indicate that POKAZATEL underestimates intensities by approximately 3\%, while our new line list agrees with the measurements.

\begin{figure}[h!]
    \centering
    \begin{subfigure}{0.50\textwidth}
        \includegraphics[width=\textwidth]{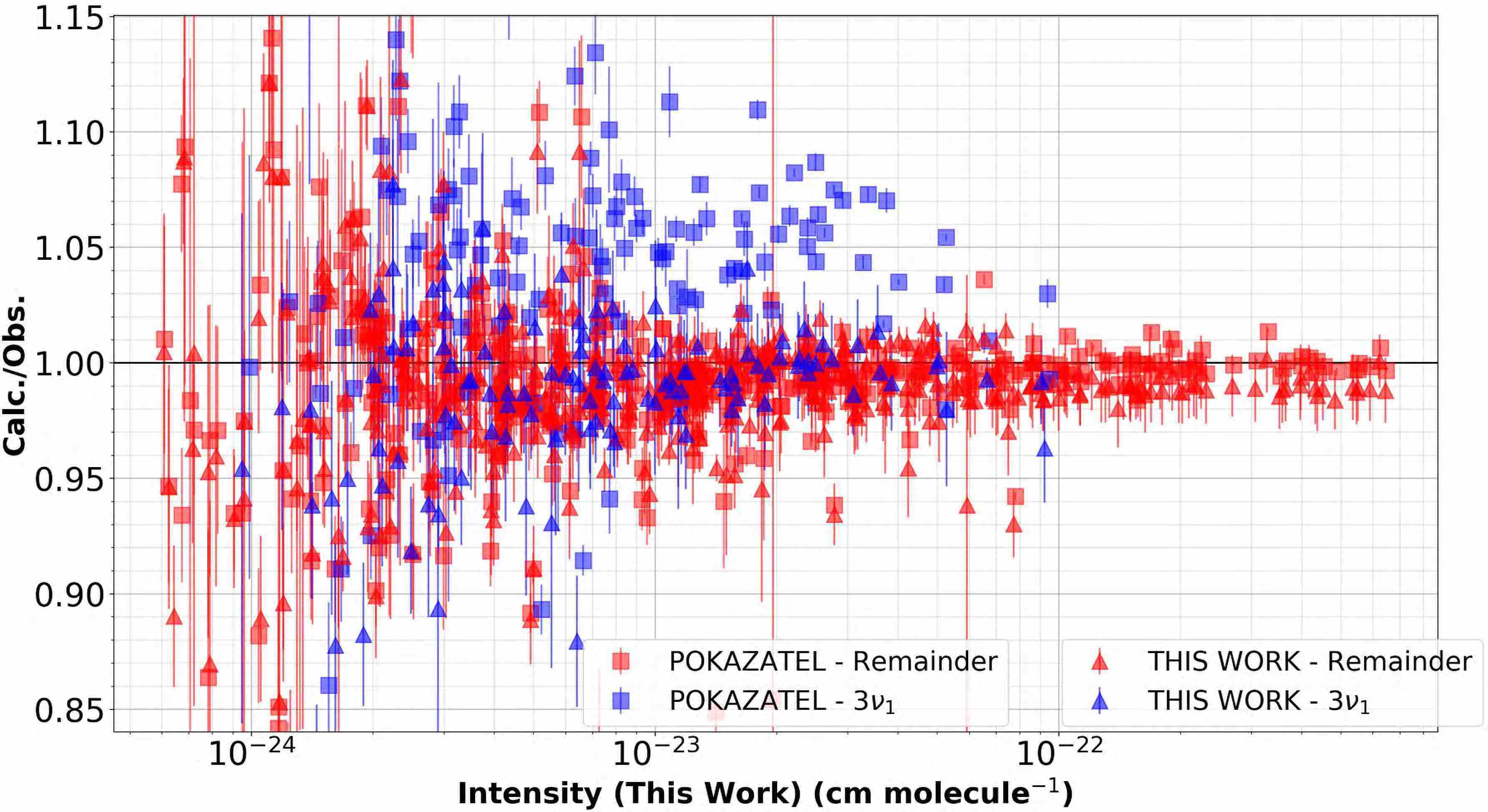}
        \caption{}
        \label{fig:6a}
    \end{subfigure}

    \begin{subfigure}{0.50\textwidth}
        \includegraphics[width=\textwidth]{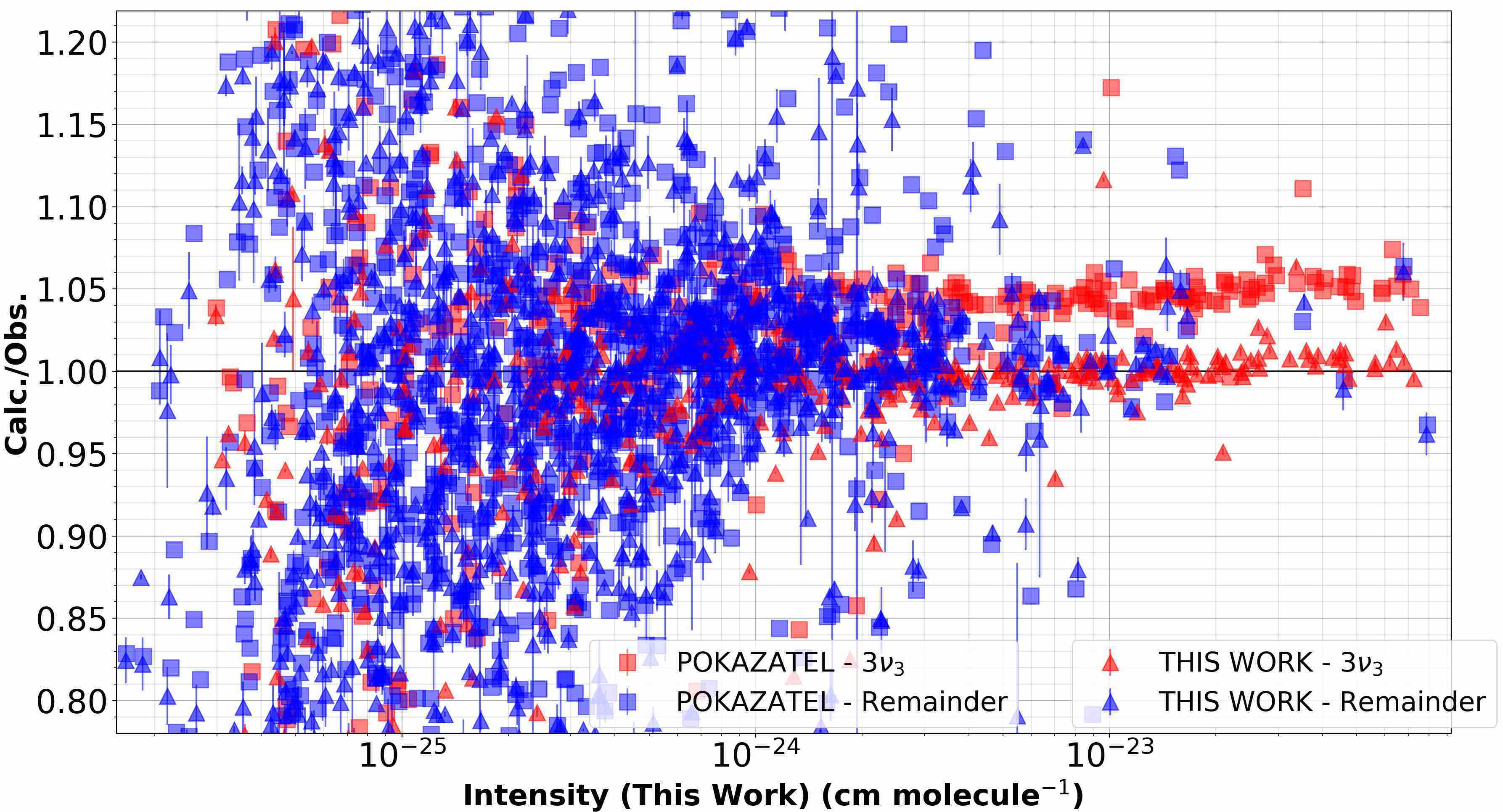}
        \caption{}
        \label{fig:6b}
    \end{subfigure}
    \caption{Near-infrared comparisons of transition intensities obtained from our new line list and POKAZATEL\cite{jt734} to the experiments of (a) Birk \textit{et al.}\cite{jt687} and (b) Brown \textit{et al.}\cite{H2O-S-15-11}. For subplot (a), error bars for POKAZATEL ratios are experimental while our work combines experimental and theoretical PES stability. For (b), the uncertainty on our work is just the PES stability. }
    \label{fig:6}
\end{figure}

In Figure \ref{fig:6a}, we compare with the measured intensities of 535 transitions provided in the study of Birk \textit{et al.}\cite{jt687} present in HITRAN2016, with frequencies covering 10000 - 11000 cm$^{-1}$. Our line list predicts intensities that are significantly closer to the reported experimental values than the line list of POKAZATEL, see Figure \ref{fig:6a}. Comparisons in the (300), (121) and (201) bands show deviations can reduce by up to 8\% with our new line list\cite{jt744}.

We matched with 1475 transitions within HITRAN2016 that originate from the work of Brown \textit{et al.}\cite{H2O-S-15-11} and plot intensity ratios in Figure \ref{fig:6b}. Brown \textit{et al.} analyzed 25 individual spectra that were measured at the National Solar Observatory, Kitt Peak and produced intensities based on these observations. For well isolated lines, uncertainties are stated to be approximately 3\%, or possibly lower in some cases. These intensities include many of the bands covered in \cite{jt687} and possess transition frequencies in the region 10240 - 11378 cm$^{-1}$. For the 3$\nu_{3}$ band, POKAZATEL systematically over-estimates the absorption by approximately of 5\%, as can seen in Figure \ref{fig:6b}.

In Table \ref{table:2}, we summarize key results for a number of vibrational bands previously discussed. For many of the bands considered, the range of intensities is extremely large, often spanning over six order of magnitude, hence we use a weighted average, see formula (1).  With the exception of the (130) band, there is overall, an improvement over POKAZATEL, particularly for transitions at the shorter wavelengths.  It is however, worth highlighting the large change in the average instability factor in the $\nu_{2}$ and $2 \nu_{2}$ bands, which increases from only $0.02$ \%, to $0.80$ \%. For water vapor, accurately modeling the bending behavior in potentials has always been difficult \cite{11BuPoZo.H2O,jt714}. This stability factor is likely to continue increasing for the higher excitations in $\nu_{2}$ and hence, the \textit{ab initio} wave-functions may reduce the accuracy of the transition intensities. 

\begin{equation}
W(S)=\left\{
\begin{array}{@{}ll@{}}
1000\frac{S}{S_{max}}, & \text{if}\ \frac{S}{S_{max}} \leq 1000 \\
1, & \text{otherwise}
\end{array}\right.
\end{equation}

There is a distinct energy gap present at 13000 cm$^{-1}$ for the H$_{2}$$^{16}$O experimental data within HITRAN2016 in the region covered by O$_{2}$ A band absorption. Incorrect intensities and/or other line parameters occurring in this region has the possibility to interfere with remote sensing experiments of O$_{2}$.

There are two different sets of experimental data in HITRAN2016 that come from the work of Tolchenov \textit{et al.}\cite{H2O-S-27-23,H2O-S-28-24}. The first \cite{H2O-S-27-23} includes those transitions with frequencies in the range 10251 - 14495 cm$^{-1}$, and the other \cite{H2O-S-28-24} continuing from 14500 - 25232 cm$^{-1}$. For the infrared-only measurements\cite{H2O-S-27-23}, we were able to compare with a total of 3911 transitions that are taken from HITRAN2016, with resulting intensity ratios displayed in Figures \ref{fig:7a}. As previously seen with comparisons to the experiment of Brown \textit{et al.}, we again observe an approximate shift of 4-5\% for the strongest lines of POKAZATEL. Campargue \textit{et al.} \cite{CAMPARGUE20082832} also recorded intracavity laser spectroscopy (ICLAS) in the narrow region of 12746 - 13558 cm$^{-1}$ (not in HITRAN2016) and we also compare to 604 of their measured intensities in Figures \ref{fig:7a}. For the few strong lines measured, we are in better agreement to experiment than POKAZATEL.

\begin{table}
	\caption{Average weighted intensity deviation in a select number of vibrational bands calculated as 100$|S_{calc}/S_{obs} - 1|$. All observed/experimental data is taken from HITRAN2016\cite{jt691}. Where an experimental error is provided, the data is from Birk \etal\cite{jt687}. TW refers to this work. }
	\hspace{-2.7cm}
	\begin{tabular}{|c|c|c|c|c|c|c|c|}
		\hline
		Band & \# Lines  & S$_{\text{min}}$ & S$_{\text{max}}$ &  Instability (\%) &Exp. Error (\%)  &TW (\%)& POKAZATEL (\%) \\
		\hline 
		100  &890     &5.8(-27)    &2.3(-20)   &0.34   &0.28    &1.51  &1.51   \\
		001  & 1019    &2.9(-26)  &2.5(-19)   & 0.04   &0.24   & 0.46 & 0.19  \\
		010  &912     &1.5(-26)    & 3.2(-19)  &0.02    &0.17   &0.55   &0.55  \\		
		020  &793    & 1.9(-26)   &2.8(-21)   &0.80    &0.47  &0.87    &1.11  \\
		130$^{a}$  &173     &1(-26)    &7.9(-24)   &1.5    & -- & 14.77 & 9.75  \\
		031$^{a}$  &142     &1(-26)    &2.8(-23)   &0.3    &--  & 2.93 &9.66   \\
		300  &120     & 9.5(-26)   &9.2(-23)   &1.65    &1.43   &2.79  &6.05   \\
		003  &151     &1.0(-24) &7.2(-23) &0.22    &--   &1.54  &4.65   \\
		201  &236     & 6.1(-25)   &6.4(-22)   &1.29    &0.99   &2.05  &1.68   \\
		121  & 116    &6.8(-25)    &1.0(-22)   &0.53    &1.37   &2.89  &2.96   \\
		\hline
	\end{tabular}
	
	$^{a}$ Deviation has been calculated by averaging the results from Campargue \etal \cite{H2O-S-68-1217} and Regalia \etal \cite{H2O-S-67-1216}.
	\label{table:2}	
\end{table}

We also consider the theoretical intensities present in HITRAN2016 that feature in this O$_{2}$ A band region; BT2\cite{jt378} and Lodi \textit{et al.}\cite{jt509} (UCL2012), see Figure \ref{fig:7b}. Comparison of our intensities to those of Lodi \textit{et al.} show clear asymmetry between the absorption features on red and blue sides of the A-band region. This is corroborated by the independent study by Geoffrey Toon (Jet Propulsion Laboratories) \cite{Toon2012} who carried out spectral fits to a ground-based solar spectrum measured with the one of the Total Carbon Column Observing Network (TCCON) \cite{Wunch2011} Fourier Transform spectrometers (FTS) from Darwin, Australia, at 87 degrees solar zenith angle on Apr 17, 2007. It was discovered that when using UCL2012 line list to retrieve water vapor in our atmosphere, the amount retrieved below and above 13000 cm$^{-1}$ are different. Figure \ref{fig:7b} echos this observation and our line list should prove to be beneficial for future works involving the O$_{2}$ A band.  

We followed this up by acquiring the source data to both works of Tolchenov \textit{et al.} \cite{H2O-S-27-23,H2O-S-28-24}, which includes measurements from 9250 - 25232 cm$^{-1}$. It is perhaps worth noting that the original experiment was carried out by Schermaul and co-workers at the Rutherford Appleton Laboratory \cite{jt269,jt270,jt285}. We have already cross compared in the one-micron region with experiments from Birk  \textit{et al.} and Brown \textit{et al.}, which include bands (300), (003), (121), (201) and (102), and the agreement between these sources and our line list is excellent; ratios are between 1.0 $\pm$ 0.1. One method of verifying/assessing the experimental data of Tolchenov \textit{et al.} is to compare our calculated intensities to these same bands but instead using the source data from Tolchenov \textit{et al.} Intensity ratios of 3260 transitions are presented in Figure \ref{fig:7c} with the respective experimental uncertainties.

\begin{figure}[H]
	\centering
	\begin{subfigure}{0.45\textwidth}
		\includegraphics[width=\textwidth]{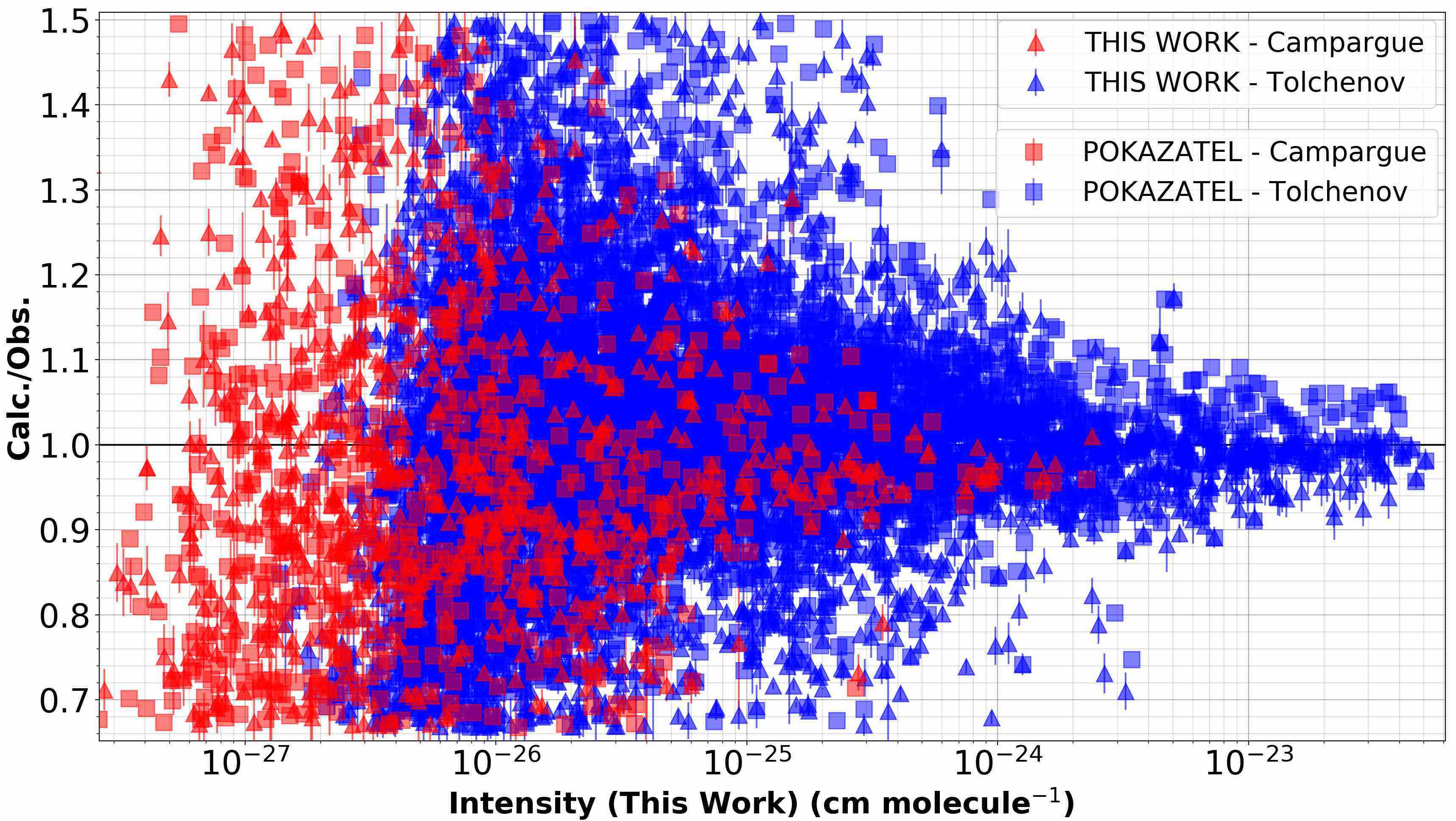}
		\caption{}
		\label{fig:7a}
	\end{subfigure}
	\\
	\begin{subfigure}{0.45\textwidth}
		\includegraphics[width=\textwidth]{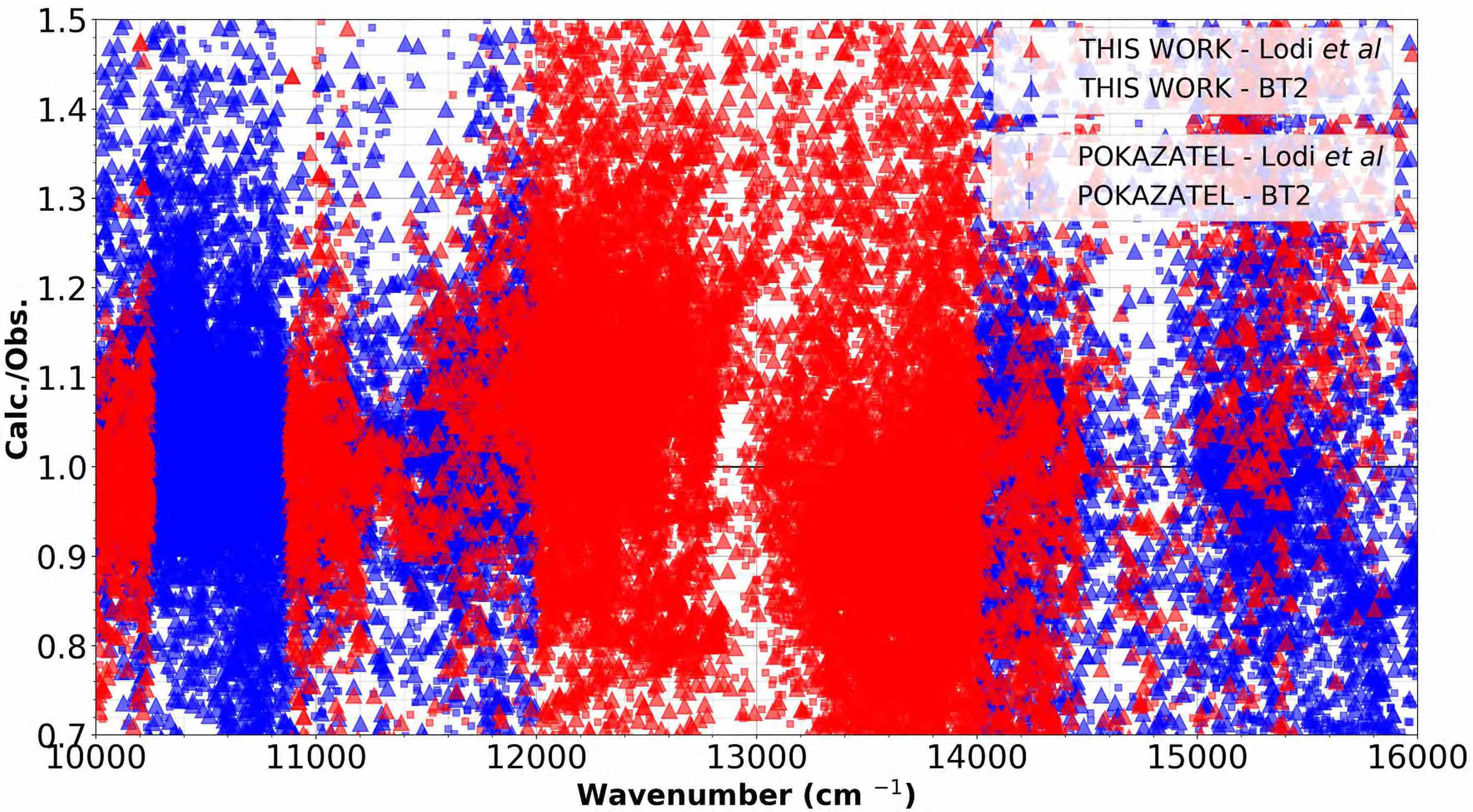}
		\caption{}
		\label{fig:7b}
	\end{subfigure}
	\\
	\begin{subfigure}{0.45\textwidth}
		\includegraphics[width=\textwidth]{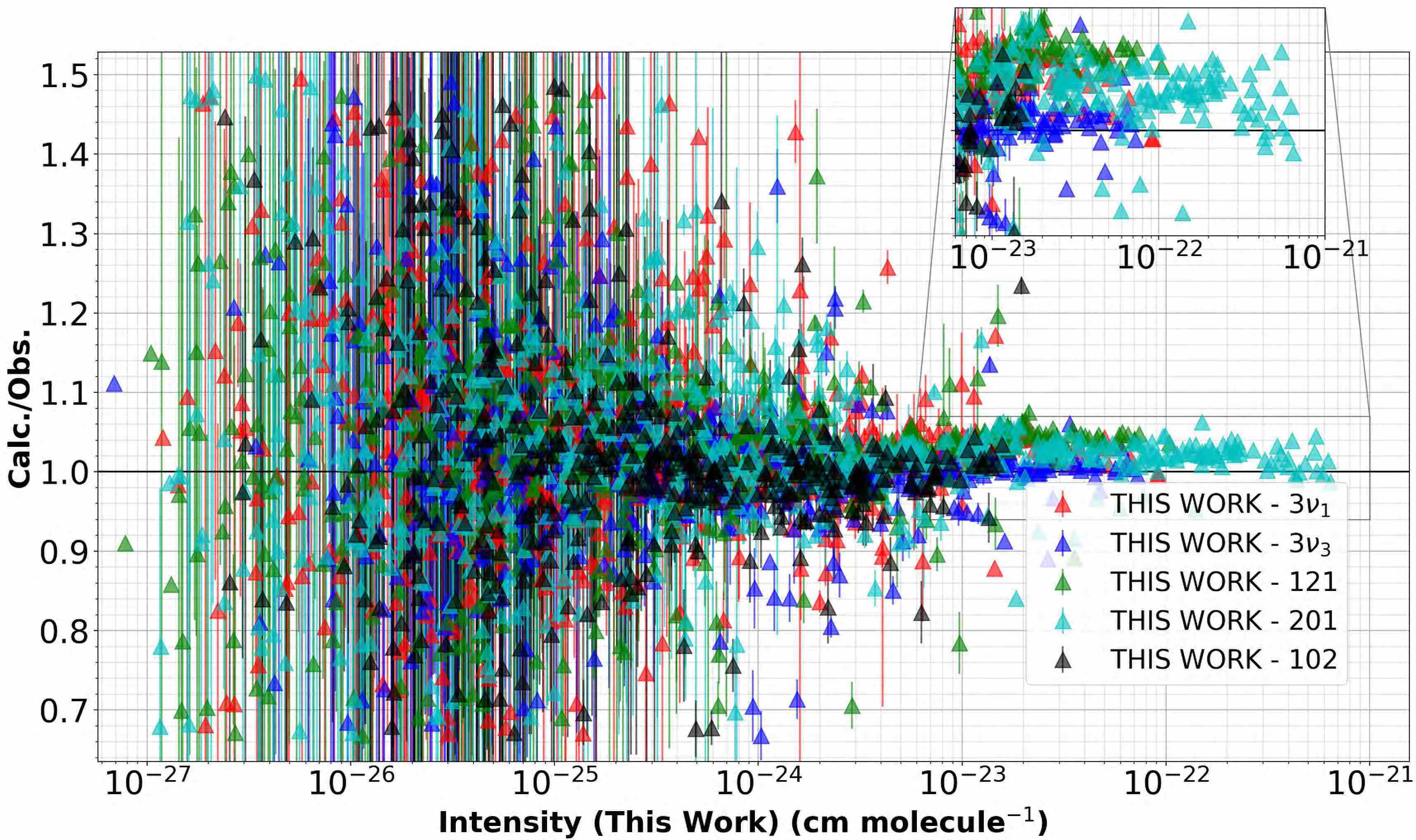}
		\caption{}
		\label{fig:7c}
	\end{subfigure}
	\caption{(a) Theoretical intensities in this work and POKAZATEL\cite{jt734} compared to the measurements of Tolchenov \textit{et al.}\cite{H2O-S-27-23, H2O-S-28-24} (in HITRAN2016) and Campargue \textit{et al.}\cite{CAMPARGUE20082832}, uncertainty is theoretical PES stability. (b) Comparison of this work against theoretical intensities of BT2\cite{jt378} and those from Lodi \textit{et al.}\cite{jt509} (UCL2012) in HITRAN2016. (c) Comparison of this work against Tolchenov \textit{et al.} data taken from the experimental source\cite{H2O-S-27-23} for bands (300), (003), (121), (201) and (102) with corresponding experimental uncertainties.}
	\label{fig:7}
\end{figure}

For transitions with intensities greater than 10$^{-24}$ cm molecule$^{-1}$, there are a large number of scattered ratios, which is concerning given these lines are relatively strong. Other comparisons in this study do not show scatter for such strong transitions, which suggests that the uncertainty on the measurements should be larger than reported. The results certainly question the use of Tolchenov \textit{et al.} data for those transitions with intensities below 10$^{-25}$ cm molecule$^{-1}$, which dominate the visible/UV spectrum, although high resolution experimental studies of this region are available \cite{jt242,jt254,jt360,jt366}.

Tolchenov \textit{et al.}\cite{H2O-S-28-24} is the only non-\textit{ab initio} source of intensity within the HITRAN2016 database for transition frequencies in the visible. Intensities in this region are in the range of 10$^{-25}$ cm molecule$^{-1}$ - 10$^{-27}$ cm molecule$^{-1}$ and considering the results we obtained in our previous comparison to those Tolchenov \textit{et al.} infrared measurements, no conclusion could be made from a line-by-line intensity comparison as ratios were scattered throughout the visible. However, for the 3569 transitions that we matched with in HITRAN2016, we observed that POKAZATEL lacks approximate labels for most upper energy states (which are simply denoted by the rigorous quantum numbers $J$, parity, ortho/para) for transitions occurring above 20000 cm$^{-1}$. 

We instead decided to calculate air broadened cross sections using the HAPI\cite{HAPI} program with the Voigt profile. However, all transitions in HITRAN2016 and HITEMP2010 contain both air and self broadening parameters, whilst POKAZATEL and our line list do not. In order to facilitate an equal comparison, equal broadening parameters should be used across all sources, hence we apply approximate broadening co-efficients to each transition that are calculated as a function of $J'$ and $J''$ \cite{HITEMP2010}.

Lampel \textit{et al.}\cite{jt645} showed through atmospheric observations, that the visible region between 20200 - 21500 cm$^{-1}$ was modeled better by HITEMP2010 than POKAZATEL. 

\begin{figure}[H]
	\centering
	\begin{subfigure}{0.49\textwidth}
		\includegraphics[width=\textwidth]{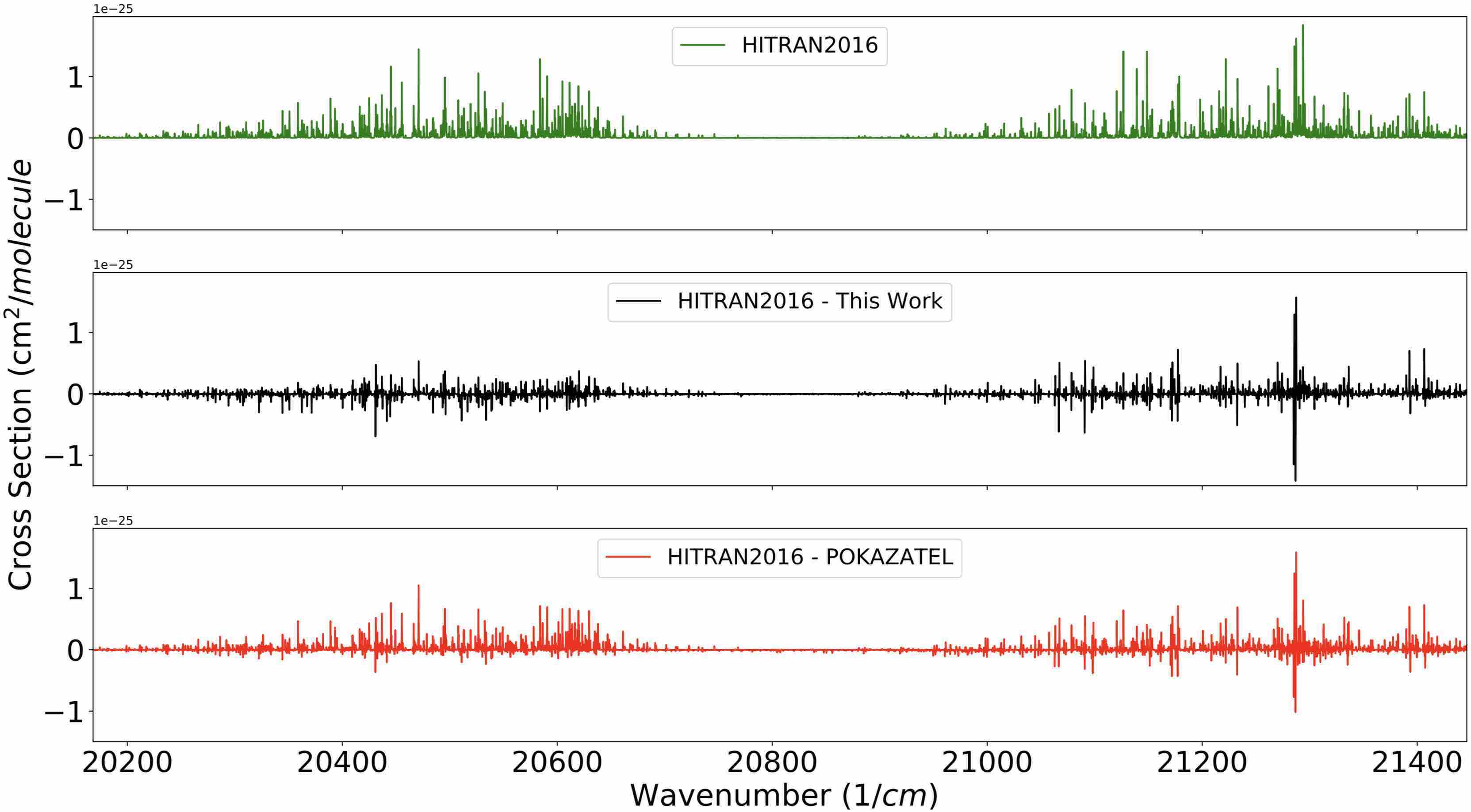}
		\caption{}
		\label{fig:8a}
	\end{subfigure}
	\begin{subfigure}{0.49\textwidth}
		\includegraphics[width=\textwidth]{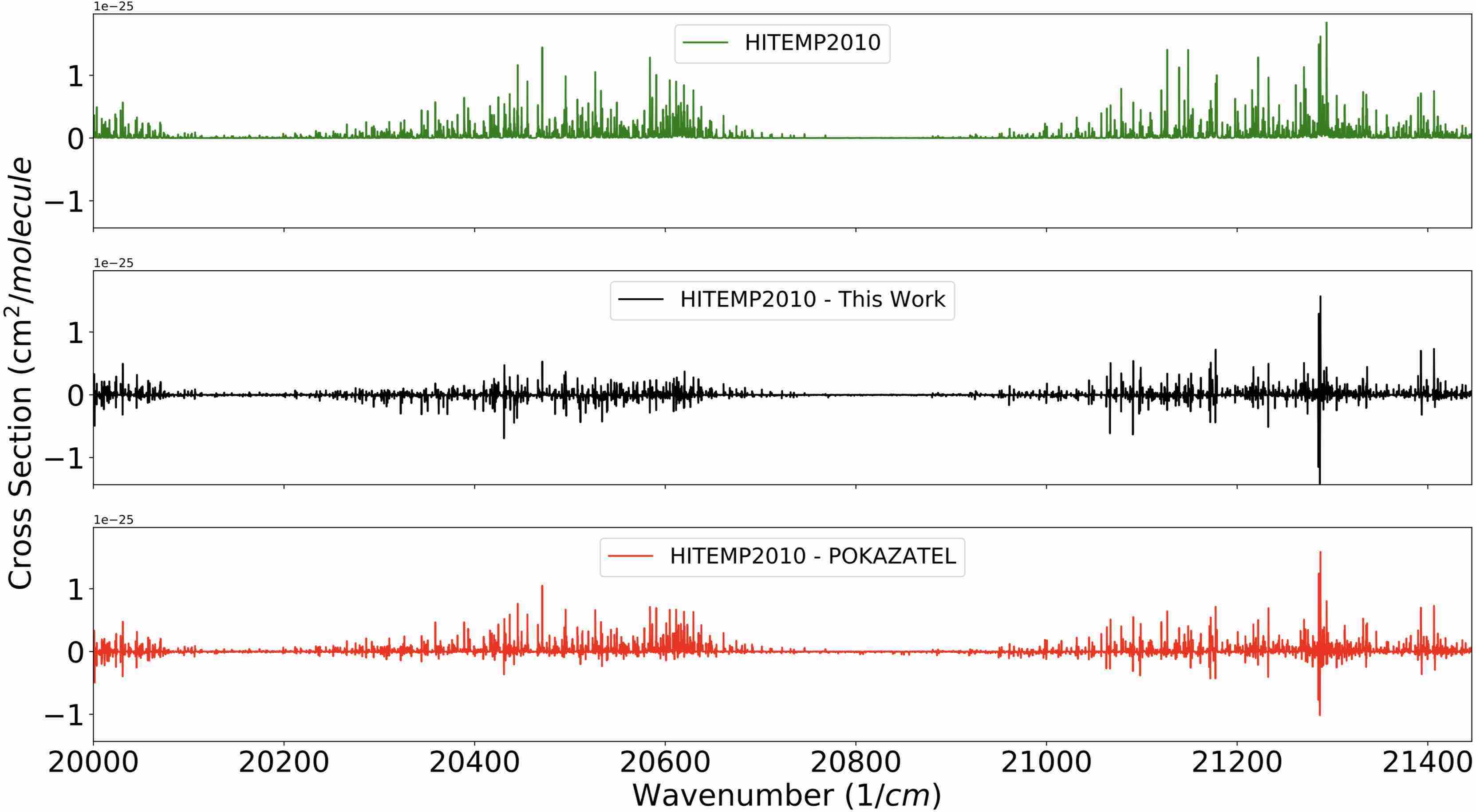}
		\caption{}
		\label{fig:8b}
	\end{subfigure}
	\\
	\begin{subfigure}{0.49\textwidth}
		\includegraphics[width=\textwidth]{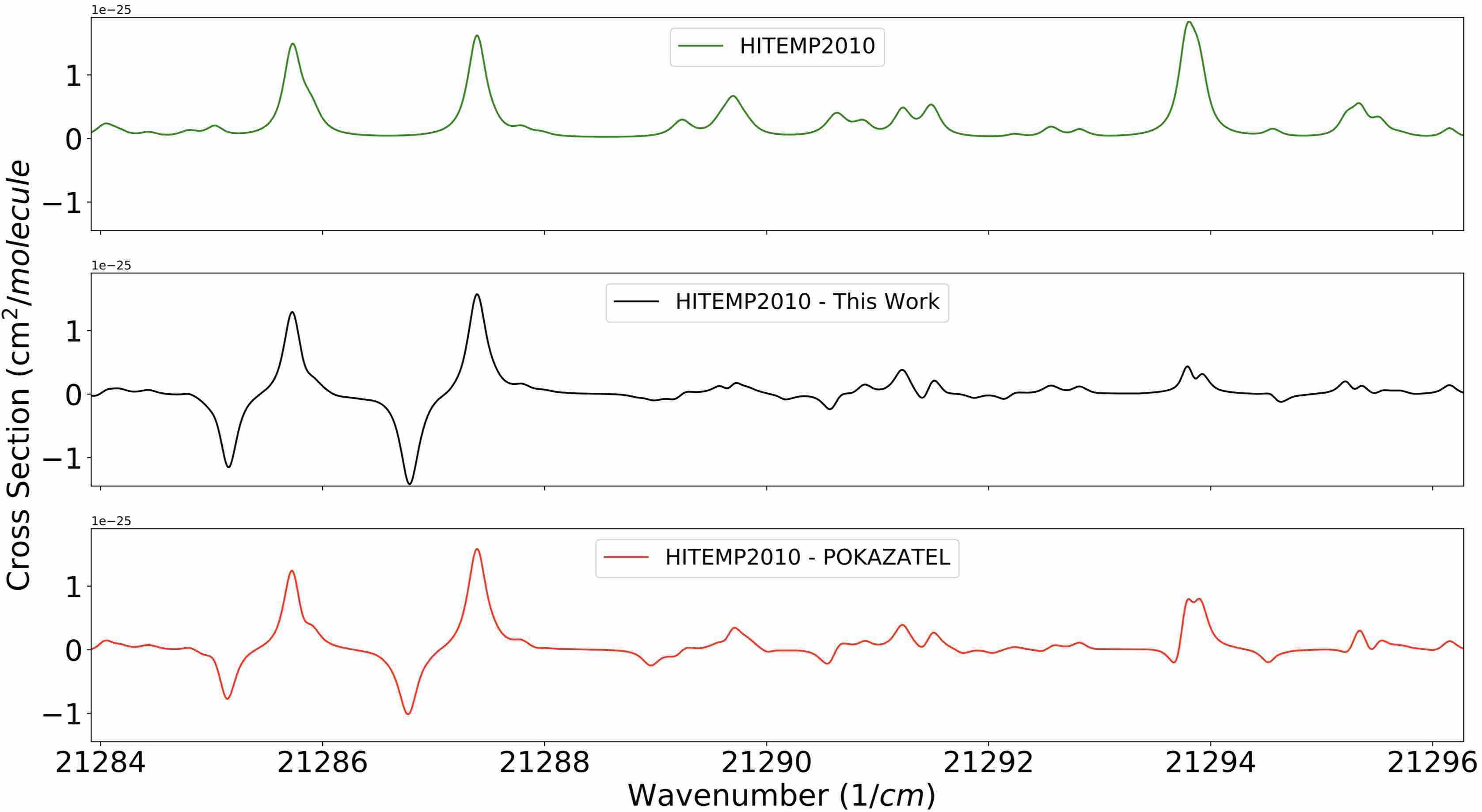}
		\caption{}
		\label{fig:8c}
	\end{subfigure}
	
	\caption{Cross-section residuals obtained from subtracting our new line list and POKAZATEL\cite{jt734} from (a) HITRAN2016, and (b), (c) HITEMP2010.}
	\label{fig:8}
\end{figure}

Figures \ref{fig:8a} and \ref{fig:8b} present residuals obtained from subtracting the theoretical cross sections produced from both POKAZATEL and our line list from those obtained using HITRAN2016 and HITEMP2010. For the first region of absorption between 20200 - 20800 cm$^{-1}$, our line list clearly possesses significantly lower residuals than POKAZATEL when comparing to both HITRAN2016 (Figure \ref{fig:8a}) and HITEMP2010 (Figure \ref{fig:8b}). The residuals are also smaller when comparing with HITEMP2010 rather than HITRAN2016.

For the absorption region located in the interval 21000 - 21500 cm$^{-1}$, our new line list also provides better agreement to both HITRAN2016 and HITEMP2010 over POKAZATEL, with the exception of line position differences occurring in our comparison to HITEMP2010, located at approximately 21285 cm$^{-1}$, seen in Figure \ref{fig:8b}. This feature is common to both theoretical line list residuals, see Figure \ref{fig:8c}. Considering this feature is not present in the comparisons to HITRAN2016 data, it is clear that some line positions in the visible spectrum of HITEMP2010 need updating. This is not surprising as HITRAN2016 has seen two updates since HITEMP2010 was introduced: HITRAN2012 and HITRAN2016.

\subsection{H$_{2}$$^{18}$O}

\begin{figure*}[h]
    \centering
    \begin{subfigure}[ht]{0.4\textwidth}
        \includegraphics[width=\textwidth]{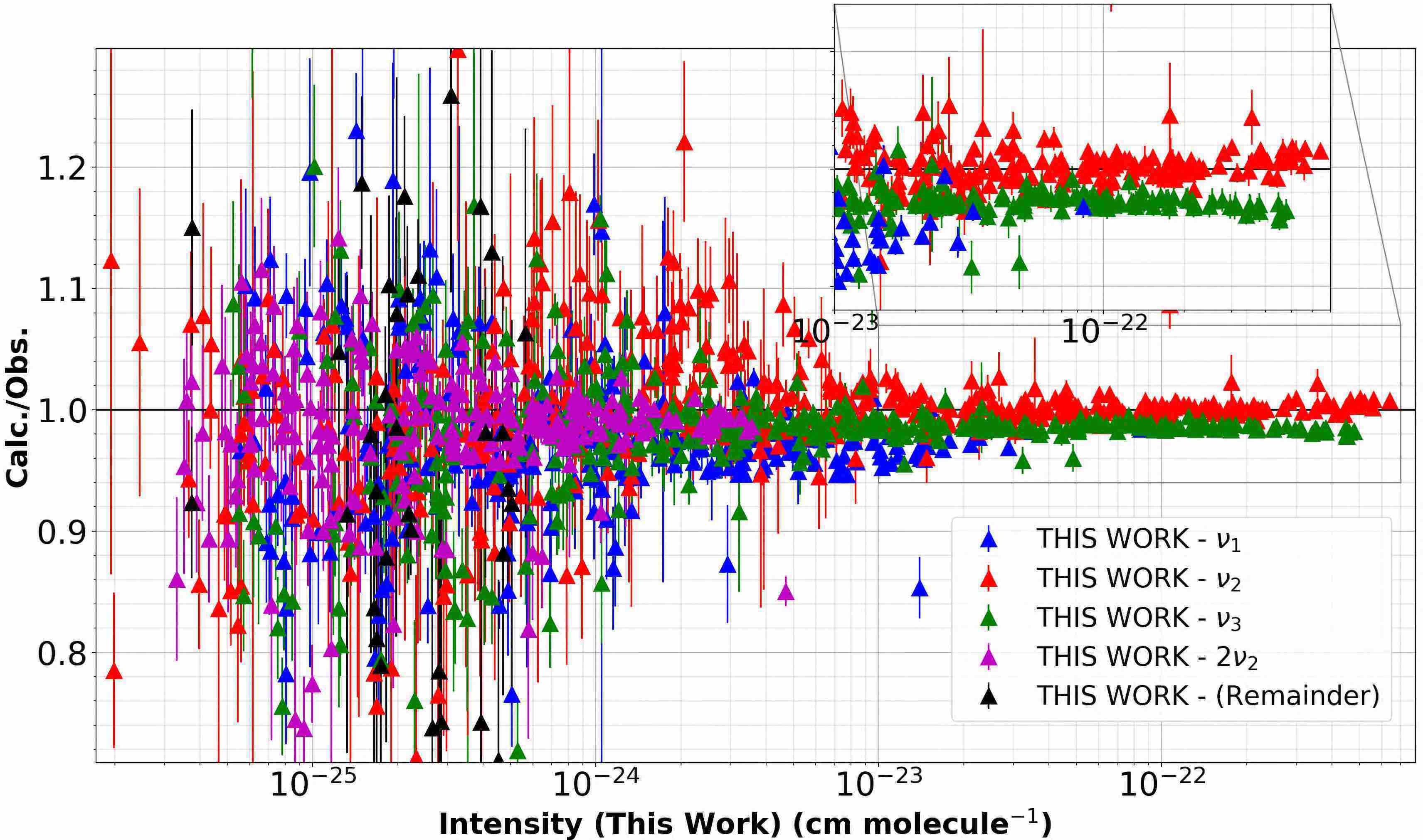}
        \caption{}
        \label{fig:9a}
    \end{subfigure}
    \begin{subfigure}[ht]{0.4\textwidth}
        \includegraphics[width=\textwidth]{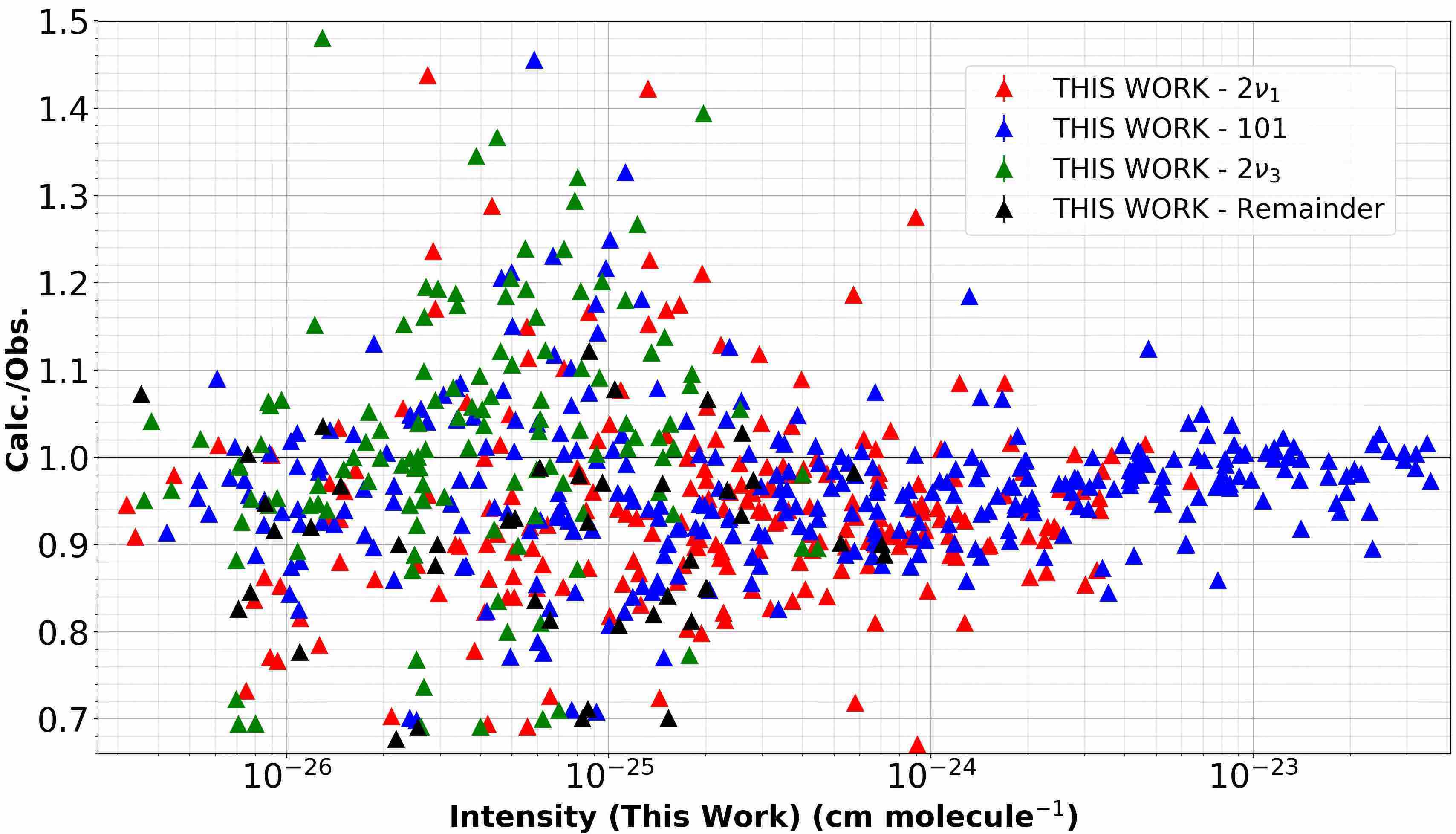}
        \caption{}
        \label{fig:9b}
    \end{subfigure}
    \\
    \begin{subfigure}[ht]{0.4\textwidth}
        \includegraphics[width=\textwidth]{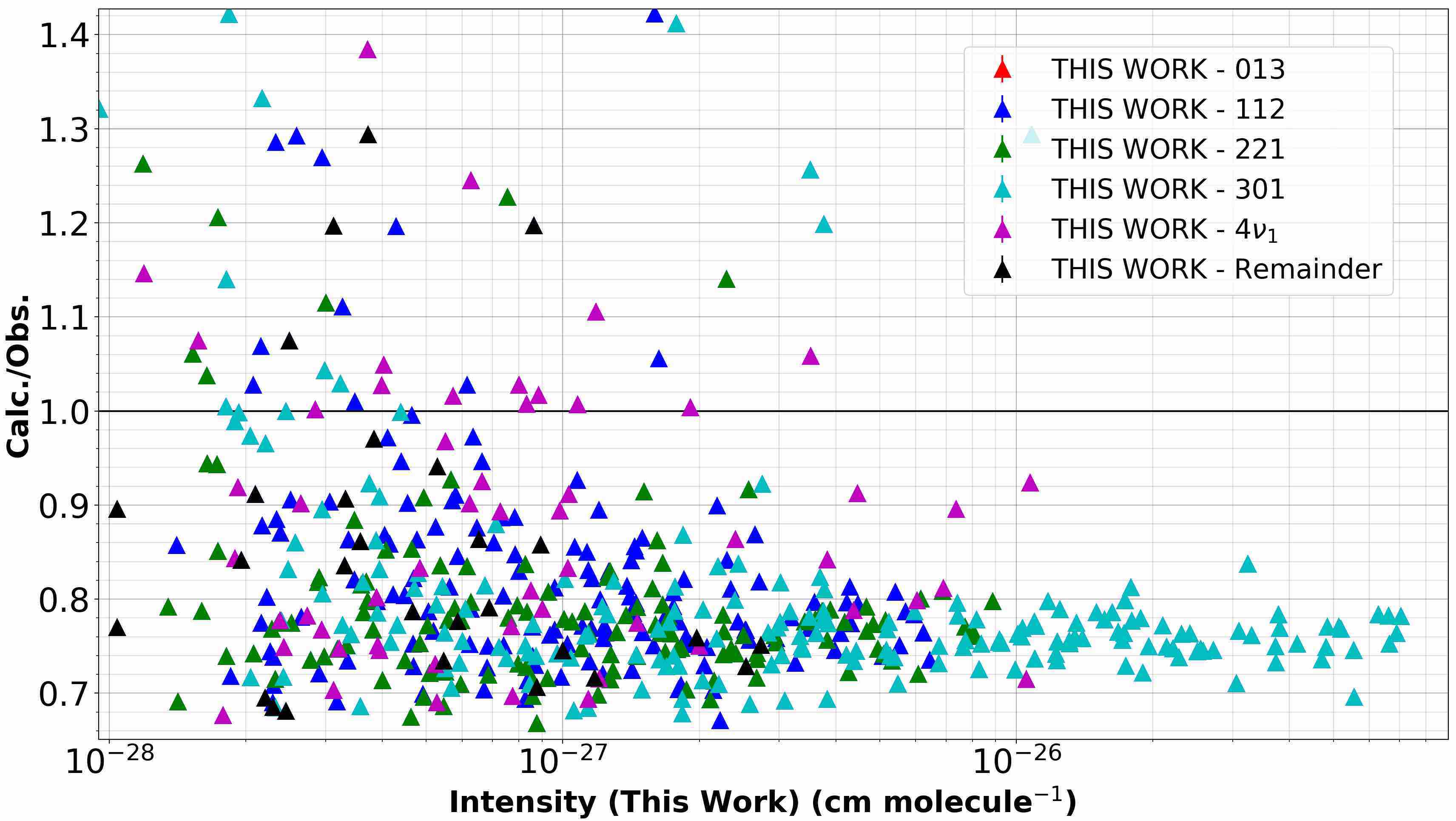}
        \caption{}
        \label{fig:9c}
    \end{subfigure}
    \begin{subfigure}[ht]{0.4\textwidth}
        \includegraphics[width=\textwidth]{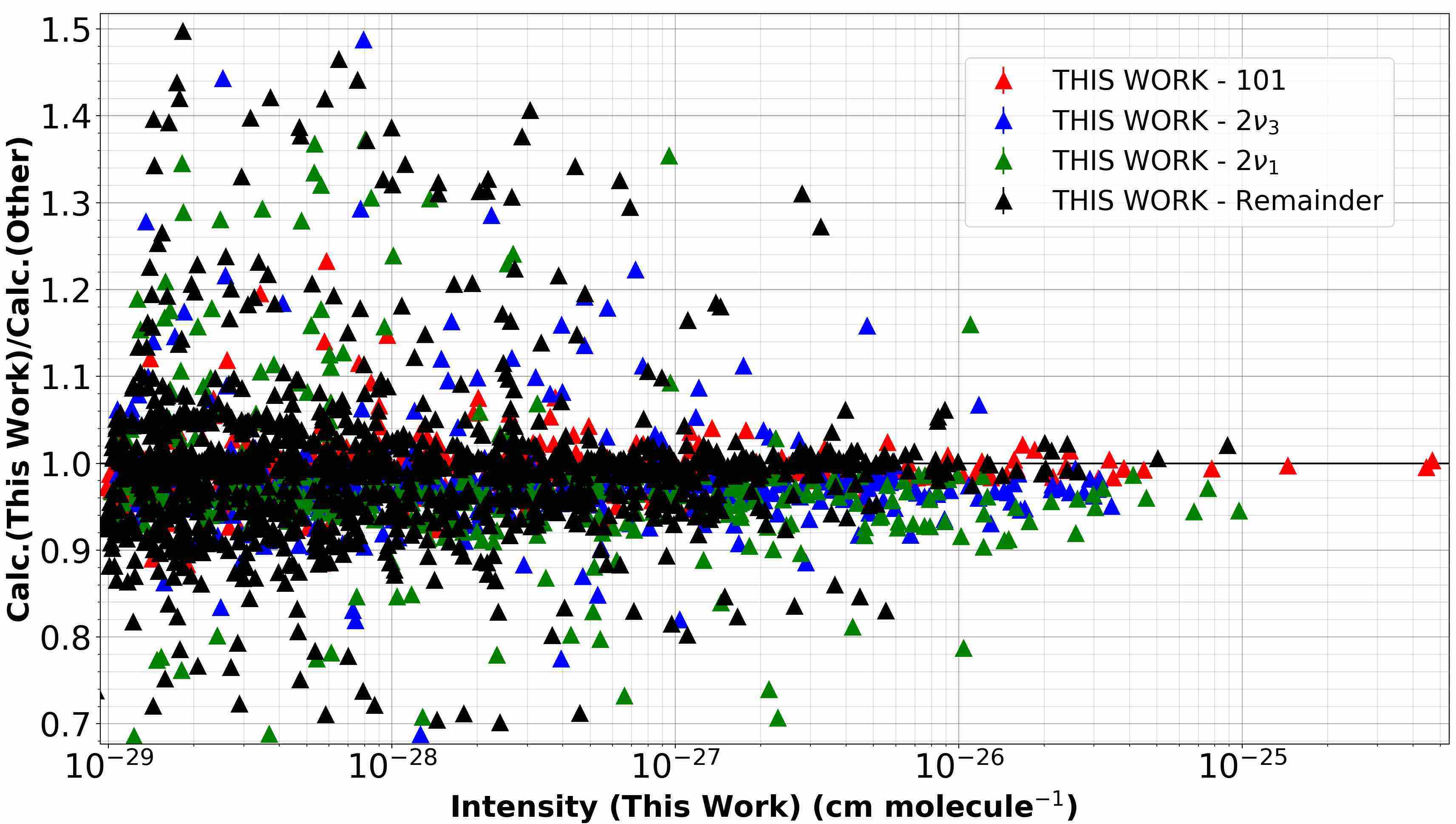}
        \caption{}
        \label{fig:9d}
    \end{subfigure}

    \caption{H$_{2}$$^{18}$O intensity ratios between this work and the experiments of (a) Loos \textit{et al.} \cite{H2O-nu-59-1235}, (b) Toth \textit{et al.}\cite{H2O-S-18-15}, (c) Tanaka \textit{et al.}\cite{jt303}, and (d) the \textit{ab initio} calculations of Partridge and Schwenke \cite{H2O-S-70-12}. Uncertainty on (a) is experimental. }
    \label{fig:9}
\end{figure*}

Calculations from Lodi and Tennyson \cite{jt552} make up the majority of the HITRAN2016 database for H$_{2}$$^{18}$O. The DMS used in their work is very similar to the DMS used in the creation of the POKAZATEL line list: they are both fit to the same electronic structure data points. Comparing our work to their H$_{2}$$^{18}$O data in HITRAN would yield near very similar results to those previously seen for POKAZATEL in H$_{2}$$^{16}$O, particularly in the IR. In the visible, one expects the differences to become more apparent.

Loos \textit{et al.} and Birk \etal \cite{BIRK1250}also measured intensities for H$_{2}$$^{18}$O in the infrared region 1260 - 3995 cm$^{-1}$ and from HITRAN2016, we compare with 1387 of these. The resulting ratios are presented in Figure \ref{fig:9a} and are very similar to those shown in Figures \ref{fig:3a} and \ref{fig:3b} where we compared to their H$_{2}$$^{16}$O measurements. This result was expected as our theoretical models for H$_{2}$$^{16}$O and H$_{2}$$^{18}$O use potentials from the same source, and the same DMS. This result simultaneously highlights the high quality of their experiment, as well as the stability of our calculations across isotopologues. 

 As expected the Lodi and Tennyson \cite{jt552} line list exhibits larger  intensity errors in the visible as was seen by Mikhailenko \etal \cite{MIKHAILENKO2018170} in their measurements of H$_{2}$$^{18}$O spectra in the 16460 - 17200 cm$^{-1}$ interval. The data of Lodi and Tennyson was found to be incomplete in this region, with several strong lines and countless weak transitions missing. On top of this, several strong transitions appeared to be consistently too strong. We have compared to HITRAN2016 in the region of interest and the same conclusions are made, see Figure \ref{fig:10}. The red boxes in Figure \ref{fig:10} are an example of missing strong lines in the database, most identified by Mikhailenko \etal 

\begin{figure}[h]
    \centering
        \includegraphics[width=0.45\textwidth]{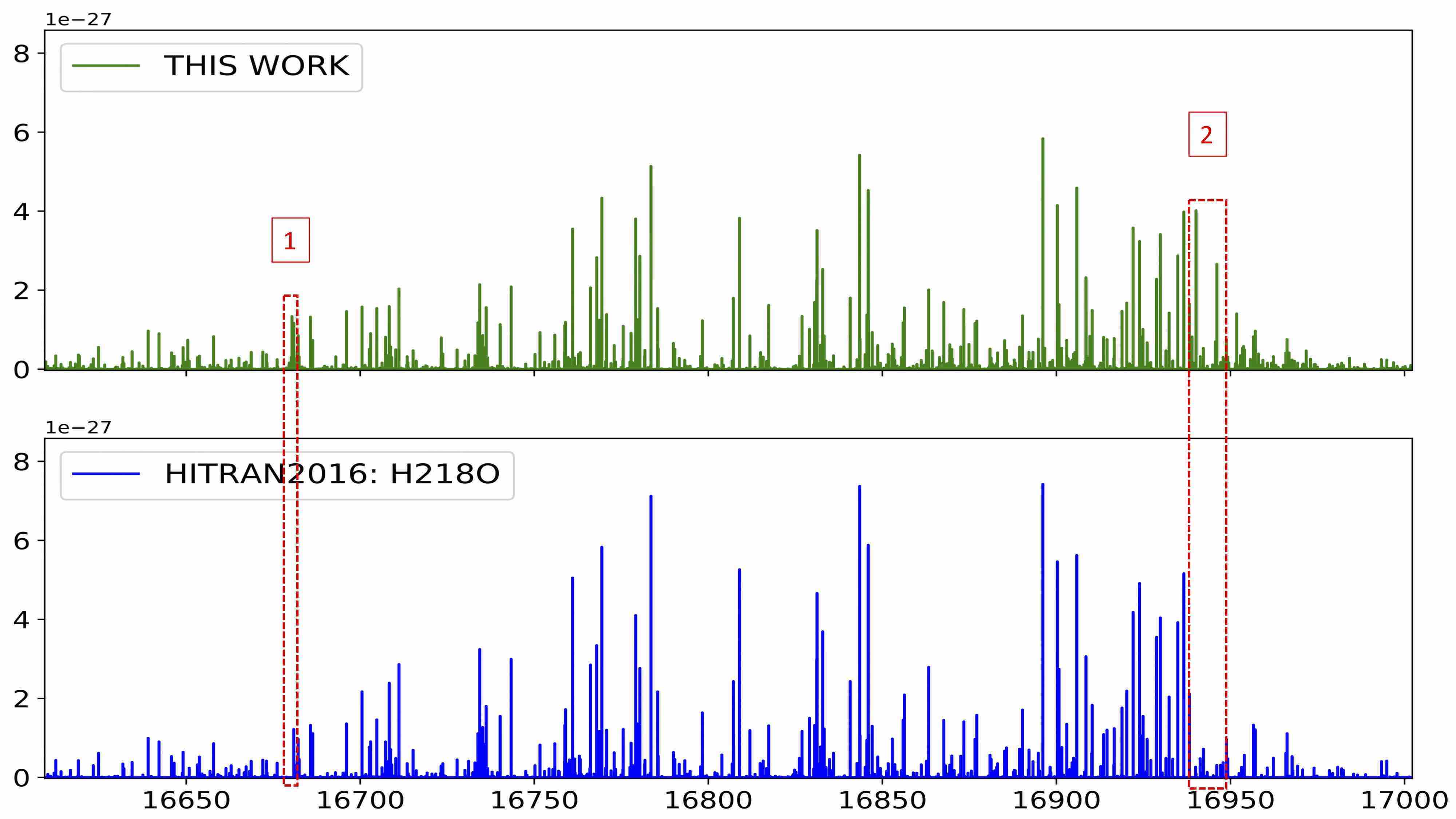}
    \caption{HITRAN2016 and our new H$_{2}$$^{18}$O spectra plotted in the visible region 16650 - 17000 cm$^{-1}$. Red boxes are an example of missing strong lines in the database.}
    \label{fig:10}
\end{figure}

The \textit{ab initio} intensities from Partridge and Schwenke \cite{H2O-S-70-12} form a large fraction of HITRAN2016 between 7000 - 8339 cm$^{-1}$. We compare with 2280 of these with intensity ratios presented in Figure \ref{fig:9b}. In general, the agreement is excellent, although there does appear to be a small systematic shift.

For H$_{2}$$^{16}$O, the agreement between our line list and the observed data of Toth \textit{et al.} present in HITRAN2016 was very good, see Figure \ref{fig:3a}. This does not appear to be the case for H$_{2}$$^{18}$O, Figure \ref{fig:9c}. We compare with 713 transitions in the region of 7000 - 7678 cm$^{-1}$, all taken from HITRAN2016. As previously explained for H$_{2}$$^{16}$O, these intensities are measured, not calculated. The intensity ratios appear significantly more scattered than those of H$_{2}$$^{16}$O, a feature that is most likely introduced from the experimental data.

Tanaka \textit{et al.}\cite{jt303} analyzed H$_{2}$$^{18}$O spectra that was previously measured at the National Solar Observatory, Kitt Peak. From HITRAN2016, we matched with 549 transitions in the region of 12404 - 14276 cm$^{-1}$ with the resultant ratios shown in Figure \ref{fig:9c}. The discrepancy is significant and appears to be systematic, with our intensities 'appearing' to be almost 22\% too small. However, we have already compared with H$_{2}$$^{16}$O intensities in this region, see Figure \ref{fig:7a}, and no 22\% shift is observed for the strongest transitions. The results suggest that there may have been an error in the evaluation of the abundances within the sample in Ref. \cite{jt303}.

We have previously compared to the calculated semi-empirical H$_{2}$$^{16}$O intensities of Coudert \textit{et al.}, Figure \ref{fig:1}, and both sets of theoretical data are in very good agreement with each other and to the $\nu_{2}$ and $\nu_{3}$ measured band intensities of Loos \textit{et al.}. However, for H$_{2}$$^{18}$O, comparisons indicate that our intensities and those of Coudert \textit{et al.}\cite{COUDERT2016130} are very  different, see Figure \ref{fig:11a}. The intensity ratios of the pure rotational transitions appears skewed, and there are systematic shifts in the intensities of both the $\nu_{2}$ and $\nu_{3}$ bands. This is not seen in our comparison to the same bands observed by Loos \textit{et al.}, hence these deviations must originate from the calculations of Coudert \textit{et al.}. We note that in our procedure a single model can be used for all isotopologues whereas the
method of Coudert \textit{et al.} requires a separate fit for each one.

We point out that the work of Coudert \textit{et al.} forms the basis of the GEISA2015 database for H$_{2}$$^{18}$O. The GEISA2015 article outlines that the data from Coudert \textit{et al.} reproduces the measured intensities of Oudot \textit{et al.}\cite{OUDOT2012859} better than other \textit{ab initio} methods available at the time. We thus compare with the measured intensities of Oudot \textit{et al.} in Figure \ref{fig:11b}. There is little agreement between our work and that of Oudot \etal and ratios in the $\nu_{2}$ band are scattered. This behavior was not seen when we compared to the experiments of Birk and Loos \etal \cite{BIRK1250,H2O-nu-59-1235} (see Figure \ref{fig:9}).  The data from \cite{BIRK1250,H2O-nu-59-1235} present in HITRAN was recently successfully validated against atmospheric spectra by Olsen \etal \cite{OLSEN2019} and this gives confidence in the accuracy of our calculation. 

\begin{figure}[H]
	\centering
	\begin{subfigure}[ht]{0.5\textwidth}
		\includegraphics[width=\textwidth]{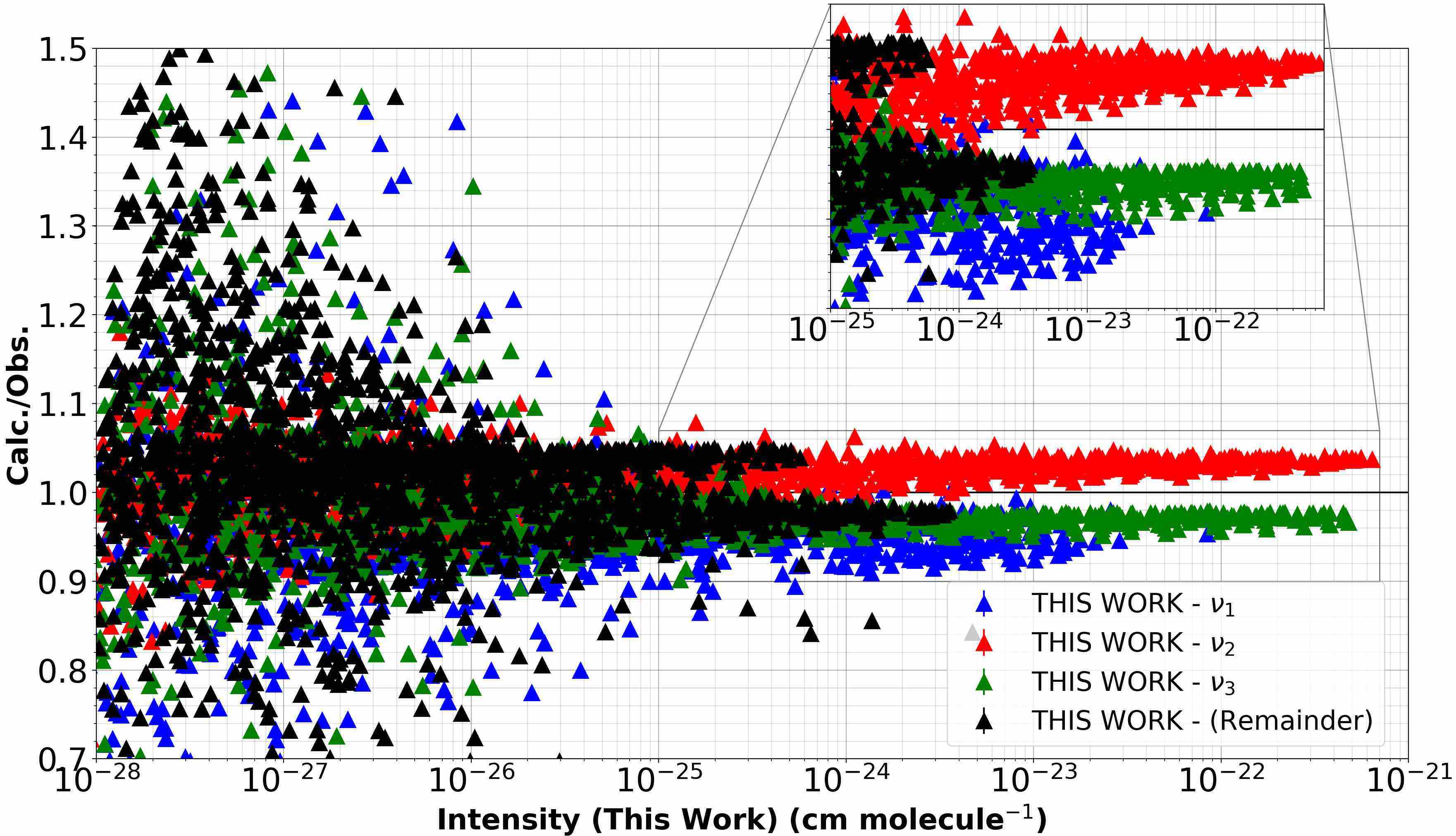}
		\caption{}
		\label{fig:11a}
	\end{subfigure}
	\begin{subfigure}[ht]{0.5\textwidth}
		\includegraphics[width=\textwidth]{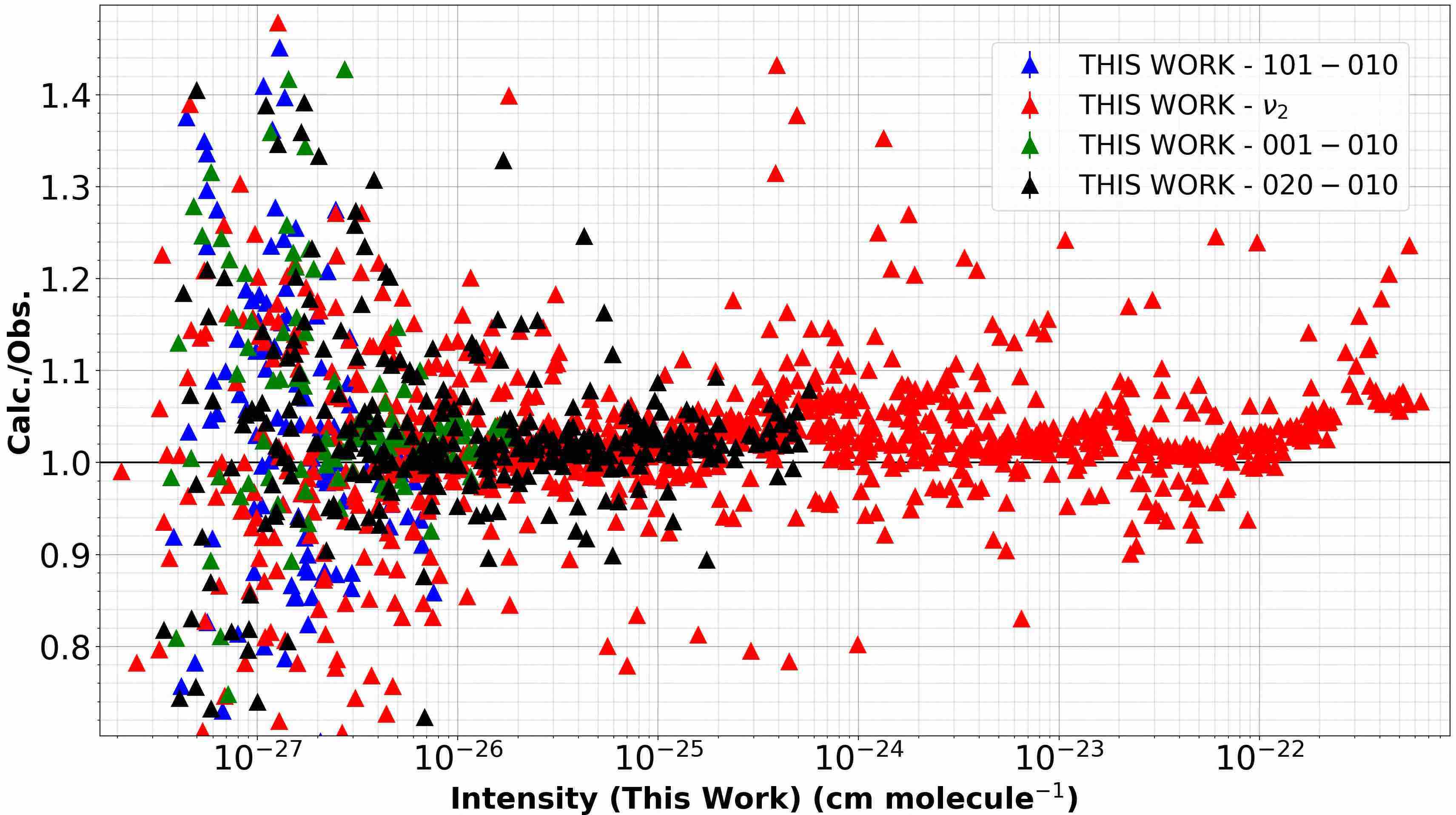}
		\caption{}
		\label{fig:11b}
	\end{subfigure}
	
	\caption{Comparison of our intensities to (a) the semi-empirical calculations of Coudert \textit{et al.}\cite{COUDERT2016130}, and (b), the experimental measurements of Oudot \textit{et al.}\cite{OUDOT2012859}.}
	\label{fig:11}
\end{figure}

 \section{Summary}
Our new line lists have been computed by combining the most accurate and computationally intensive global dipole moment surface for water vapor with highly accurate semi-empirical potential energy surfaces for each isotopologue, which predict H$_{2}$$^{16}$O energy levels to within 0.022 cm$^{-1}$ for those lower than 26000 cm$^{-1}$. Where possible, we have replaced our calculated energy levels for each line list with the semi-empirical levels predicted from MARVEL. Where this was done, labels were also placed on each state: $K_{a}$, $K_{c}$, $\nu_{1}$, $\nu_{2}$ and $\nu_{3}$. For our H$_{2}$$^{16}$O line list, we calculate a 'potential stability' for many transition intensities based on the interchange of two potential energy surfaces. This can be significant for unstable transitions.

Comparisons have been made against eighteen different sources across both H$_{2}$$^{16}$O and H$_{2}$$^{18}$O that encompass measured transition intensities from the far-infrared to the visible, which total 24890 observations. In the microwave, we also compare to calculated semi-empirical intensities. 

The majority of the measured data was obtained from HITRAN2016, as it is frequently used in both the characterization of terrestrial atmospheres and in comparisons with new experiments and is known to provide accurate results, hence it is the most logical source to compare with. 

In the microwave, the agreement between our theoretical intensities to those from the semi-empirical calculations of Coudert \textit{et al.} are excellent for the strongest lines for H$_{2}$$^{16}$O. For H$_{2}$$^{18}$O, systematic shifts are observed in  $\nu_{2}$ and $\nu_{3}$ bands, of which these differences are attributed to the semi-empirical calculations. Errors are also observed in semi-empirical pure rotational H$_{2}$$^{18}$O intensities, which feature in GEISA2015. 

In the far-infrared, intensities from our new H$_{2}$$^{16}$O line list and the line list of Polyansky \textit{et al.} (POKAZATEL) show excellent agreement to two different experimental sources, one of which is not in HITRAN2016.

For the mid-infrared spectrum of H$_{2}$$^{16}$O, seven individual sources provide intensities that cover 7000 - 8339 cm$^{-1}$. The POKAZATEL line list over-estimates the absorption of the (031) band by approximately 10\%, seen in two of the seven studies. POKAZATEL also underestimates the absorption of the 2$\nu_{3}$ band by 3\%. Our new line list accurately models these two bands and intensities are within experimental uncertainties.

For H$_{2}$$^{18}$O, there is an increased amount of scatter in the intensity ratios obtained from comparing with the Toth \textit{et al.} data within HITRAN2016 compared to what is obtained when considering Toth \textit{et al.} H$_{2}$$^{16}$O intensities. It is likely that this difference is not of theoretical origin. Our new H$_{2}$$^{18}$O line list also shows excellent agreement with the \textit{ab initio} intensities from Partridge and Schwenke, although it appears there is a small systematic shift in all intensities. 

We compared with three different experiments in HITRAN2016 that provide H$_{2}$$^{16}$O intensities measurements in the near-infrared spectrum. Our new line shows significant improvement in this region, up to 5-8\% in many bands, notably (300), (003), (121), (102) and (201). For H$_{2}$$^{18}$O, the near-infrared region in HITRAN2016 includes intensities coming from Tanaka \textit{et al.} Intensity ratios show a substantial 25\% offset when compared against our line list. This shift is not observed when we compared to a different experiment for H$_{2}$$^{16}$O that covers the same frequency range. 

Water vapor absorption in the molecular Oxygen A band region which covers approximately 12000 - 13000 cm$^{-1}$ currently remains widely disputed. Comparisons of our new intensity calculations to two independent experiments yields unconvincing results; scatter is a dominant feature in all but the strongest transitions.

Within the visible section of HITRAN2016, we could not compare our calculated H$_{2}$$^{16}$O intensities to the experimental measurements of Tolchenov \textit{et al.} on a line-by-line basis as we question the accuracy of their measurements below approximately 10$^{-24}$ cm molecule$^{-1}$. Hence, we instead calculated cross-sections.

A recent atmospheric study in the visible \cite{jt645} concluded that HITEMP2010 provided better agreement to observation than the POKAZATEL line list. When comparing cross sections from our new H$_{2}$$^{16}$O line list to both HITRAN2016 and HITEMP2010 cross sections, our line list shows smaller residuals than what is obtained when using POKAZATEL. 

Line position differences were however observed in our comparisons to the HITEMP2010 database that are not replicated through comparisons against HITRAN2016. This is due to continued improvements in semi-empirical potentials which provide better line positions in HITRAN2016.

As we expected, very similar intensity ratios are obtained when comparing our H$_{2}$$^{16}$O and H$_{2}$$^{18}$O line lists to the same experimental source of Loos \textit{et al.} The line lists were calculated using potentials from the same source, as well as the same dipole moment surface, hence all line lists should yield similar results across the same bands. 

Across the entire spectrum, our new H$_{2}$$^{16}$O and  H$_{2}$$^{18}$O line lists are in excellent agreement with many experiments and offer significant improvements in the prediction of numerous band intensities when compared to many H$_{2}$$^{16,18}$O line lists. It should prove to be a useful resource for the future updates of transition intensities within the HITRAN database, from the microwave to the visible. For intensities in the (130) band, and other bands, where there is a systematic offset against two different experiments, these \textit{ab initio} calculations would need to be scaled. Theoretical work is currently on-going to further improve the visible spectrum H$_{2}$$^{16}$O. Each line list is available in the supplementary material.

\section*{Acknowledgments}

HITRAN and HITEMP databases are supported through the NASA Aura (NNX17AI78G) and PDART (NNX16AG51G) grants. Nikolai F. Zobov and Aleksandra A. Kyuberis acknowledge support by State project IAP RAS No0035-2019-0016.

\bibliographystyle{elsarticle-num}



\end{document}